\documentclass{article}
\usepackage{aaspp4}
\usepackage{epsfig}
\usepackage{astrobib}
\usepackage{hyperref}

\def \beginfig          {\begin{figure}}
\def \endfig            {\end{figure}}

\def \begineq           {\begin{equation}}
\def \endeq             {\end{equation}}

\def \d {\partial}

\def\gtorder{\mathrel{\raise.3ex\hbox{$>$}\mkern-14mu
             \lower0.6ex\hbox{$\sim$}}}
\def\ltorder{\mathrel{\raise.3ex\hbox{$<$}\mkern-14mu
             \lower0.6ex\hbox{$\sim$}}}

\def \lsim {\ltorder}
\def \gsim {\gtorder}
\def \hide#1{}
\def \half {{1\over 2}}

\def \max {{\rm max}}

\def\br {{\bf r}}

\def\bz {{\bf z}}

\def \beginfig          {\begin{figure}}
\def \endfig            {\end{figure}}
\def \figcap            {\caption}

\def \begineq           {\begin{equation}}
\def \endeq             {\end{equation}}

\def \obs	{{\rm obs}}
\def \fobs 	{f_{\rm o}}
\def \fint 	{f}
\def \fs 	{f_{\rm s}}
\def \fcrit 	{f_{\rm crit}}

\def \gdag	{g^\dagger}

\def \d {\partial}

\def \Maij {M_{\alpha i j}}

\def \Malm {M_{\alpha l m}}

\def \P {{\cal P}}
\def \R {{\cal R}}

\def \Peff {{\overline P}}

\newcommand{\astrophref}[1]{\href{http://xxx.lanl.gov/abs/astro-ph/#1}{\tt{http://xxx.lanl.gov/abs/astro-ph/#1}}}

\begin{document}

 
\title{A New Shear Estimator for Weak Lensing Observations}
\author{Nick Kaiser}
\affil{
Institute for Astronomy, University of Hawaii \\
2680 Woodlawn Drive, Honolulu, HI 96822 \\
e-mail: \href{mailto:kaiser@hawaii.edu}{\tt kaiser@hawaii.edu} \\
web-site: \href{http://www-nk.ifa.hawaii.edu/~kaiser}{\tt http://www-nk.ifa.hawaii.edu/$\sim$kaiser}
}

\begin{abstract}
We present a new shear estimator for weak lensing observations which
properly accounts for the effects of a realistic point spread function (PSF).
Images of faint galaxies are subject to gravitational shearing
followed by smearing with the instrumental and/or atmospheric
PSF.  We construct a `finite resolution shear operator' which when
applied to an observed image has the same effect as a gravitational
shear applied prior to smearing.  This operator allows one to 
calibrate essentially any shear estimator.  We then specialize to
the case of weighted second moment shear estimators.  We compute the
shear polarizability which gives the response of an individual galaxy's
polarization to a gravitational shear.  We then compute the response of the
population of galaxies, and thereby construct an optimal weighting scheme
for combining shear estimates from galaxies of various shapes, luminosities 
and sizes.
We define a figure of merit --- an inverse shear variance per unit
solid angle --- which characterizes the quality of image data for shear
measurement.  The new method is tested with simulated image data.
We discuss the correction for anisotropy of the PSF and
propose a new technique involving measuring shapes from images which have
been convolved with a re-circularizing PSF.  
We draw attention to a hitherto ignored noise related bias
and show how this can be analyzed and corrected for.
The analysis here draws heavily
on the properties of real PSF's and we include as an appendix a brief review,
highlighting those aspects which are relevant for weak
lensing.
\end{abstract}


\section{Introduction}
\label{sec:introduction}

Weak lensing provides a probe of the dark matter distribution on
a range of scales from galaxy halos, through clusters of
galaxies, to large-scale-structure (see e.g.~the recent review of 
\citeN{mellier99}, and references therein).
In the weak lensing or thin lens approximation, 
the effect of gravitational lensing on the image of
a distant object is a mapping of the surface brightness:
\begineq
\label{eq:lensmapping}
f'(r_i) = f((\delta_{ij} - \psi_{ij}) r_j)
\endeq
where the 2-vector $r_i$ is the angular position on the sky
measured relative to the center of the image, 
$f$ is the intrinsic surface brightness that would be 
seen in the absence of lensing, and the symmetric `distortion
tensor' $\psi_{ij}$ is an integral along the line of sight of the transverse
second derivatives of the gravitational potential $\Phi$ \cite{gunn67}.  
In an open cosmology,
for instance, the distortion for an object at conformal distance $\omega_s$ 
can be written as
\begineq
\label{eq:psifromphi}
\psi_{lm} = 2 \int d \omega {\sinh \omega \sinh (\omega_s - \omega) \over 
\sinh \omega_s} \partial_l \partial_m \Phi
\endeq  
(\citeNP{barkana96}; \citeNP{k98})
where $\partial_i \equiv \partial / \partial x_i$ and $x_i = \theta_i
\sinh \omega$ with $\theta_i$ being a 2-component Cartesian vector in the
plane of the sky and where the potential is 
related to the density contrast by $\nabla^2 \Phi = 4 \pi G \delta \rho$,
the Laplacian here being take with respect to proper spatial coordinates.
Equation (\ref{eq:psifromphi}) can be generalized to deal
with sources at a range of distances, and with either accurately
known redshifts or partial
redshift information from broadband colors.

The distortion will, in general, change both the shapes and sizes,
and hence luminosities, of distant objects.  Any component of the
distortion which is coherent over large
scales --- larger than the typical angular separation of background galaxies ---
is therefore potentially observable as a relative modulation of the counts of objects
or as a statistical anisotropy of the galaxy shapes, and this allows one
to constrain the fluctuations in the total density $\delta \rho$ on the corresponding
scales.
Here we will focus on analysis of shape anisotropy, or `image shear', 
though the methodology is readily
extendible to include the effects of magnification.  

While the effect of a shear on the sky surface brightness (\ref{eq:lensmapping})
is rather simply stated, no completely satisfactory method for
estimating the distortion has yet emerged.  Perhaps ideally one would
attack this problem using likelihood; that is, one would ask: what is the probability
to observe a given set of background galaxy images given that they are drawn from
some statistically isotropic unlensed parent distribution, 
as a function of the parameters
$\psi_{lm}$.  
Unfortunately, this does not
seem to be a particularly tractable problem.  Also, the problem is further complicated by
the finite resolution of real observations, and by noise in the images.  Instead, what
has been done is to adopt some plausible shape statistic --- typically some kind of
central second moment --- and then compute how this responds to a gravitational
shear.  We will now review the 
various 
different shear estimators that have been proposed, how they relate
to one another, and what their limitations are.  

\subsection{Projection Matrices and the Shear Operator}

As a preliminary, we now
introduce some mathematical formalism which will simplify the analysis.  
Symmetric $2 \times 2$ tensors like $\psi_{lm}$ feature 
prominently in what follows.  Such tensors have 3 real degrees of freedom.  For instance,
it is conventional to
parameterize the three real degrees of freedom of $\psi_{lm}$ by the 
triplet comprising the convergence
$\kappa$ and the shear $\gamma_\alpha$, $\alpha = 1,\;2$ with
\begineq
\psi_{ij} = \left[\matrix{
\kappa + \gamma_1 & \gamma_2 \cr
\gamma_2 & \kappa - \gamma_1
}\right]
\endeq
A simple way to convert between triplet and tensor components is to use
the three constant $2 \times 2$ `projection
matrices': 
\begineq
\label{eq:Mdefinition}
M_{0ij} = 
\left[
\matrix{
1 & 0 \cr
0 & 1
}
\right] 
\quad\quad\quad\quad
M_{1ij} = 
\left[
\matrix{
1 & 0 \cr
0 & -1
}
\right] 
\quad\quad\quad\quad
M_{2ij} = 
\left[
\matrix{
0 & 1 \cr
1 & 0
}
\right] 
\endeq
The symmetrized products of pairs of these matrices 
$[M_A M_B] \equiv \half (M_{A l m} M_{B m n} + M_{A n m} M_{B m l})$
have multiplication table
\begineq
[M_A M_B]
= 
\left[\matrix{
M_0 & M_1 & M_2 \cr
M_1 & M_0 & 0   \cr
M_2 & 0   & M_0
}\right]
=
\delta_{AB} M_0 + 
(\delta_{A0} \delta_{B\alpha} + \delta_{A\alpha}\delta_{B0}) M_\alpha
\endeq
from which follow the useful identities that the contractions of
products and triple products are
\begin{eqnarray}
\label{eq:traceMAMB}
M_{Alm} M_{Bml} & 	= & 2 \delta_{AB} \\
\label{eq:traceMAMBMC}
M_{Alm} M_{Bmn} M_{Cnl} & = & 
2(\delta_{BC} \delta_{A0} + \delta_{AC} \delta_{B0} + \delta_{AB} \delta_{C0}).
\end{eqnarray}   
Any symmetric tensor $t_{lm}$ can be written as a linear combination of 
projection matrices with coefficients $t_A$, that is, $t_{lm} = t_A M_{Alm}$,
and using (\ref{eq:traceMAMB}) we have
$M_{Blm}t_{lm} = M_{Blm} M_{Alm} t_A = 2 \delta_{AB} t_A$, so 
$t_A = {1\over 2} M_{Alm} t_{lm}$.
In this language the convergence and the shear are the three components of
the triplet representation of the distortion tensor:
$\kappa = {1 \over 2} M_{0lm} \psi_{lm} = \psi_0$ and
$\gamma_\alpha = {1 \over 2} M_{\alpha lm} \psi_{lm} = \psi_\alpha$, $\alpha = 1,2$.
We will adopt the convention that upper case Roman indices range over $0,1,2$
while lower case Greek symbols range over $1,2$ and that repeated indices are
to be summed over.  
A two component `polar' 
$t_\alpha$ transforms under rotations as $t_\alpha \rightarrow R_{\alpha \beta}(2 \theta) t_\beta$
while $t_0$ transforms as a scalar under rotations.
An alternative and widely used formalism \cite{sef92}
is to regard $\gamma_1,\;\gamma_2$ as the real
and imaginary parts of a complex shear, but we shall not adopt that
approach here.

It is also convenient to define a `shear operator' $S_\gamma$,
which generates the mapping (\ref{eq:lensmapping}).  
\begineq
f' = S_\gamma f
\endeq
At linear order in $\gamma$, which will be valid for sufficiently weak
shear,  one can perform a first order Taylor expansion
of the RHS of (\ref{eq:lensmapping}) and $S_\gamma$ becomes the differential operator
\begineq
\label{eq:linearshearoperator}
S_\gamma = 1 - \gamma_\alpha \Maij r_i \partial_j
\endeq
where $\d_j \equiv \partial / \partial r_j$.
This operator is rather similar to a rotation operator.  An important question
is what is the domain of validity of (\ref{eq:linearshearoperator}).
The answer depends on the content of the image to which it is applied.
For an image containing information only at spatial frequencies below some upper limit $k_{\rm max}$
this will be a good approximation provided
$r \ll 1 / (\gamma k_\max)$, so for finite shear
(\ref{eq:linearshearoperator}) will not apply for
spatial frequencies $\gsim 1 / (\gamma r)$. The limit here is that
combination of frequency and distance from the origin that
a shear of strength $\gamma$ corresponds to a local translation of
about an inverse wavenumber. 
We would however expect
(\ref{eq:linearshearoperator}) to correctly describe the effect of shear on the
low frequency behavior of an image, in the sense that if applied to an
image which has had the high frequency information removed, then the result
will be essentially identical to the low-frequency content of an
exactly sheared image.

\subsection{Second Moment Shear Estimators for Perfect Seeing}

A pure shear will cause a circular object to become elliptical, 
and will change the ellipticities of non-circular objects.
Following \citeN{vjt83} a
natural choice of statistic to measure such a distortion
is the second central moment or quadrupole moment
\begineq
\label{eq:secondmomentdefinition}
q_{lm} = \int d^2 r \; r_l r_m f(r) 
\endeq
where we will assume that the total flux, which is
unaffected by a pure shear, has been normalized such that
$\int d^2 r \; f(r) = 1$, and
where the origin of coordinates has been taken such that the dipole 
moment $d_l = \int d^2 r \; r_l f(r)$ vanishes. 
The triplet coefficients of the symmetric matrix $q_{lm}$ are
\begineq
\label{eq:pdefinition}
q_A = {1 \over 2} M_{Aij} q_{ij}
=\left[\begin{matrix}{
(q_{xx} + q_{yy})/2 \cr
(q_{xx} - q_{yy})/2 \cr
q_{xy}
}\end{matrix}\right]
\endeq
The first component $q_0$ is a measure of the size, or area,
of the object, while the latter two $q_\alpha$ are a measure of the eccentricity of the
object --- they vanish for a circular object --- which
we will refer to as the `polarization'.
We can compute how the quadrupole moment is affected by a shear
by applying (\ref{eq:linearshearoperator}) to $f$ in (\ref{eq:secondmomentdefinition}) to find
\begineq
q'_{lm} = q_{lm} - \gamma_\alpha \Maij 
\int d^2 r \; r_l r_m r_i \d_j f
= q_{lm} + 2 \gamma_\alpha M_{\alpha l i} q_{im}
\endeq
where we have integrated by parts and invoked the symmetry 
($M_{\alpha i j} = M_{\alpha j i}$) and
tracelessness ($M_{\alpha i i} = 0$) of the matrices $M_1$, $M_2$.
With $q_{im} = M_{Bim} q_B$ etc., and using(\ref{eq:traceMAMBMC}), we find
\begineq
\delta q_A = \half M_{\alpha l m} (q'_{lm} - q_{lm}) = 
\gamma_\alpha M_{Alm} M_{\alpha l i} M_{Bim} q_B = 
2 \gamma_\beta(\delta_{A0} q_\beta + \delta_{A\beta} q_0)
\endeq
or equivalently
\begineq
\label{eq:deltap}
\begin{matrix}{
q'_\alpha = q_\alpha + 2 \gamma_\alpha q_0 \cr
q'_0 = q_0 + 2 \gamma_\alpha q_\alpha
}\end{matrix} 
\endeq
A weak gravitational shear therefore causes a change in the
the
polarization
$\delta q_\alpha = 2 q_0 \gamma_\alpha$, 
which is proportional to the area $q_0$,
and it also induces a change in the area
$\delta q_0$ which is proportional to the eccentricity.
For intrinsically randomly oriented galaxies the average intrinsic
polarization
$\langle q_\alpha \rangle$ vanishes by symmetry, so
$\langle q'_\alpha \rangle = 2 \langle q_0 \rangle \gamma_\alpha $
and $\langle q'_0 \rangle = \langle q_0 \rangle$ and therefore
\begineq
\label{eq:simpleshearestimate}
\gamma_\alpha = \langle q'_\alpha \rangle / 2 \langle q'_0 \rangle
\endeq
which
which one can use to estimate the shear on a patch of the sky
by replacing the averaging operator $\langle \ldots \rangle$ by 
a summation: $\sum \ldots / N$.

This type of shear estimator was first introduced and used by \citeN{vjt83}
in their search for large scale shear.  
The averages of $q_\alpha$ and $q_0$ here are heavily
weighted towards larger galaxies, which
is not ideal.
More commonly, what has been done (e.g.~\citeNP{tvjm84}) is to
normalize the polarization by the trace of the second moments
and define the `ellipticity vector'
\begineq
\label{eq:naiveshearestimator}
e_\alpha = q_\alpha / q_0
\endeq
which depends only on the galaxy shape, and
whose expectation value is
\begineq
\langle e'_\alpha \rangle = \left\langle {q_\alpha + 2 \gamma_\alpha q_0 \over
q_0 + \gamma_\beta q_\beta} \right\rangle 
\simeq 2 \gamma_\alpha -2 \gamma_\beta \langle q_\alpha q_\beta / q_0^2 \rangle
= 2 \gamma_\alpha (1 - \langle e^2 \rangle / 2)
\endeq
and where we have kept only terms up to linear order in the shear
(\citeNP{ksb95}, hereafter KSB).

An alternative \cite{bm95} is to normalize
the second moments by the square-root of the determinant of $q_{lm}$
rather than by the trace:
\begineq
\label{eq:BMshearestimator}
e^{\rm BM}_\alpha = {q_\alpha \over \sqrt{|q_{lm}|}}
= {q_\alpha \over \sqrt{q_0^2 - q_\alpha q_\alpha}}
\endeq
which, again at first order in $\gamma$, has expectation value
\begineq
\langle e^{\rm BM}_\alpha \rangle = 2 \gamma_\alpha 
\left\langle \sqrt{1 + ( e^{\rm BM}_\alpha )^2}\right\rangle.
\endeq
Yet another possibility is
to use only the information in the position angle $\theta =
(\tan^{-1} q_2 / q_1) / 2$ \cite{kochanek90}, or equivalently in the
unit shear vector $|e_\alpha|$,
and there are numerous other similar statistics one could use. 

These formulae for the response of the polarization statistics should be
used with caution.  While formally correct, the averages
here must be taken in an unweighted manner over all galaxies.
This is generally neither possible 
nor particularly
desirable, as there are detection limits, and one would
ideally like to estimate the shear as some optimally weighted combination of
polarizations of galaxies of different fluxes and sizes, and these simple
relations no longer hold.  
We will return to this later.  First, however, let us consider the 
effect on these estimators of a finite point spread function,
which is well illustrated by the simple estimator 
(\ref{eq:simpleshearestimate}).

\subsection{Second Moment Shear Estimators for Finite Seeing}

In real 
observations, some or all of atmospheric turbulence,
optical aberrations; aperture size; guiding or registration
errors; atmospheric dispersion; finite pixel size;
scattering etc.~will combine to
give an observed surface brightness
\begineq
\label{eq:fobsdefinition}
\fobs(r) = \int d^2 r'\; g(r') f(r - r') \equiv g \otimes f
\endeq
where $g(r)$ is the point spread function (PSF). The
combined effect of gravitational shearing followed by instrumental
and atmospheric seeing is the transformation
\begineq
\label{eq:fobsprimedefinition}
\fobs' = g \otimes S_\gamma f
\endeq

Circular seeing will tend to reduce the ellipticity
while departures from circularity will introduce an artificial
polarization.  
To make accurate shear measurements we need to correct for the latter and calibrate the
former.
For
estimators like 
(\ref{eq:simpleshearestimate}),
which are computed from moments $q_{lm}$
as defined in (\ref{eq:secondmomentdefinition}),
the effect of the PSF is rather simple
since, as noted by \citeN{vjt83},
the second central moment of a convolution of two 
normalized functions is just the sum of the
second central moments:
\begineq
\label{eq:momentsum}
q_{lm}(\fobs) = q_{lm}(f) + q_{lm}(g)
\endeq
and this additive property is shared by the independent
components $q_A$. Thus
one can recover the second moments $q_A(f)$ that would be measured
by a large perfect telescope in space from the observed moments $q_A(\fobs)$
simply by subtracting the moments of the PSF 
$q_A(f) = q_A(\fobs) - q_A(g)$.
These may be measured from shapes of foreground
stars in the image, or, in the case of diffraction limited
seeing, computed from the
telescope design.
In terms of observed moments,
the shear estimator 
(\ref{eq:simpleshearestimate}) becomes
\begineq
\label{eq:calibratedsimpleestimator}
\hat \gamma_\alpha = {\langle q_\alpha - q^\ast_\alpha \rangle \over
2 \langle q_0 - q_0^\ast \rangle }
\endeq
where superscript $\ast$ denotes the value for a stellar object.
This
procedure --- using (\ref{eq:momentsum}) to correct the 
measured moments of the PSF
--- simply and rather elegantly compensates for both the circularizing
and distorting effects of realistic seeing.

\subsection{Weighted Moment Shear Estimators}

Unfortunately,  while very simple to analyze,
the shear estimator constructed from the second moments
as defined in (\ref{eq:secondmomentdefinition})
is not at all useful when applied to real data.  For one thing, 
photon counting noise introduces an uncertainty
in the moments which diverges as the square of the radius out
to which one integrates, and the effect of neighboring objects
will similarly grossly corrupt the signal.  There is also the
problem that for the kinds of PSFs that arise in real telescopes,
the second moment is not well defined.
To obtain a practical estimator, it is necessary to
truncate the integral in (\ref{eq:secondmomentdefinition}).  This can be done in
various ways; the approach implemented in the FOCAS software
package \cite{jt81} and also the Sextractor package
\cite{ba96} is to
truncate the integral at some isophotal threshold $\fcrit$ and
compute moments of the non-linear function 
$F(\fobs) = \Theta(\fcrit - \fobs) \fobs$,
where $\Theta(f)$ is the Heaviside function. In fact, isophotal moments
are most commonly computed from an image which has been
smoothed with some kernel --- usually an approximate model of the PSF itself ---
in order that the isophotal boundary be well defined.
An alternative, as advocated by \citeN{bm95} and by KSB, 
is to limit the range of integration with a 
user supplied weight function $w(r)$ and define
\begineq
\label{eq:weightedmomentdefinition}
q_{lm} \propto \int d^2 r\; w(r) r_l r_m \fobs(r).
\endeq
Another possibility is to define a polarization vector in
terms of the second derivatives of a smoothed image
$f_s = w \otimes f$ as
\begineq
q_{lm} \propto \d_l \d_m \int  f_s
\endeq
evaluated at the peak of $f_s$.
For a Gaussian weight function $w(r)$, however, this is essentially equivalent to
the weighted second moment (\ref{eq:weightedmomentdefinition}).
As we shall see, the weighted moment statistics, 
being linear in the surface brightness, offer significant advantages
over the isophotal threshold method,
but in either case, the simple relation (\ref{eq:momentsum})
between observed and intrinsic second moments  
no longer holds, and compensating for the effects
of the PSF becomes considerably more complicated than 
equation (\ref{eq:calibratedsimpleestimator}).

A partial solution to this problem has been offered by KSB, who 
computed the response of shear estimators like (\ref{eq:naiveshearestimator})
to an anisotropy of the PSF under that assumption that this can be modeled
as the convolution of a circular PSF $g_{\rm circ}(r)$ with some compact 
but possibly highly anisotropic function $k(r)$.  This would be a reasonable
approximation for example for the case of atmospheric seeing
in the presence of small amplitude guiding errors. 
They found that such a PSF anisotropy would introduce an artificial ellipticity
$\delta e_\alpha = P^{\rm sm}_{\alpha\beta}(\fobs) p_\beta$
where $p_\beta \equiv M_{\beta l m} \int d^2r \; r_l r_m k(r)$ 
is the unweighted polarization of $k(r)$ 
(though see \citeN{hfks98} who corrected a minor error in the analysis).
The `smear polarizability' $P_{\alpha\beta}^{\rm sm}$
is a combination of weighted moments of $\fobs$, 
and is essentially a measure of the inverse area
of the object. An interesting feature of this analysis is that only the
second moment of the convolving kernel appears here; all other details
of $k(r)$ are irrelevant, and it is relatively straightforward to determine
$p_\alpha$ from observed stars, set up a model for how this two-vector field
$p_\alpha(r)$ varies across the field, and then correct, at linear order
at least, the ellipticities to what would have been measured by a telescope
with a perfectly circular PSF.

KSB also computed the response of the ellipticity to a shear applied to $\fobs$
{\sl after\/} smearing with the PSF and found
$\delta e_\alpha = P^{\gamma}_{\alpha\beta}(\fobs) \gamma_\beta(q)$
where $P^\gamma_{\alpha\beta}$ is another combination of moments of $\fobs$ .
This of course is
not what one wants, as one really needs the response to a shear applied
before smearing with the PSF, and they suggested that one use
deep images from the Hubble Space Telescope to empirically deduce a
correction for finite seeing.
\hide{
There are various ways one can proceed.  One possible
approach for calibrating
ground based observations is to take , subject these to an
artificial shear and then degrade the observations to simulate ground-based
seeing, as was done by \citeN{ksb95}. 
By measuring how the shapes respond to the 
input shear one can empirically determine a response function or
`shear polarizability' 
$P^\gamma = \d \langle q \rangle / \d \gamma$
which gives the desired calibration factor.
The main limitation of this approach is the paucity of
deep HST fields, which makes it hard to obtain a truly representative
sample of objects.
Another problem is that this procedure cannot
be used to calibrate HST observations themselves, and additionally,
there may be slight problems in calibrating ground-based observations
due to differences in the filter transmission functions.
}
A related approach, suggested by  \citeN{wcf96a}
is to iteratively deconvolve
the images, apply a shear and then re-convolve and
again use the change in the polarization with applied shear to
calibrate the relation between $\gamma$ and the polarization measured
from the original images.

\citeN{lk97} have used a somewhat different approach.  They 
noted that the real operation (\ref{eq:fobsprimedefinition}) can be written as
$g \otimes S_\gamma f = S_\gamma((S_\gamma^{-1} g) \otimes f)$
i.e.~applying a shear before smearing is equivalent to smearing with
an anti-sheared PSF $S_\gamma^{-1} g$ and then shearing.  Now if the PSF is Gaussian, 
applying a weak shear to it is
precisely equivalent to smearing it with another 
small but anisotropic Gaussian, so the effect of this can therefore be computed using
the smear polarizability of KSB, and
it then follows that for a nearly circular Gaussian PSF, the operation 
(\ref{eq:fobsprimedefinition})
will cause a response
\begineq
e_\alpha \rightarrow e_\alpha' = e_\alpha + \delta e_\alpha
= e_\alpha + P^\gamma_{\alpha\beta} \gamma_\beta
+ P^{\rm sm}_{\alpha\beta} p_\beta(S_\gamma^{-1} g)
\endeq
where $p_\beta(S_\gamma^{-1} g)$ is the unweighted polarization of the
anti-sheared PSF and is of first order in $\gamma$.
One can most easily infer the value of $p_\beta(S_\gamma^{-1} g)$ from the
values of the shear and smear polarizabilities for
a stellar object: shearing a point source has no
effect, so for a star we must have 
$p_\alpha = - (P^\gamma(g) / P^{\rm sm}(g)) \gamma_\alpha$
where we have suppressed the indices on the stellar polarizabilities
--- for a nearly circular PSF these are approximately diagonal ---
and hence the net effect of a real shear in this approximation is
\begineq
\label{eq:lkpolarisability}
\delta e_\alpha = (P_{\alpha\beta}^\gamma(\fobs) -
(P^\gamma(g) / P^{\rm sm}(g)) P_{\alpha\beta}^{\rm sm}(\fobs)) \gamma_\beta.
\endeq
This is nice, as it expressed the linear response of the 
polarization $e_\alpha$
to a shear entirely
in terms of the observables $\fobs$ and $g$, but it
rests somewhat shakily on the assumption that shearing
a realistic PSF can indeed be modeled as smearing
it with some compact kernel.  For a Gaussian PSF this is exact,
but this is a rather special, and unfortunately unrealistic, case.

One indication that (\ref{eq:lkpolarisability}) cannot apply
for a general PSF comes from considering the factor 
$(P^\gamma(g) / P^{\rm sm}(g))$. At no point have we specified
the scale of the weighting function $w(r)$, so this
factor must be invariant to choice of scale length.  For a Gaussian
PSF this is indeed the case, but for a PSF generated by
atmospheric turbulence for instance this is not the case,
and (\ref{eq:lkpolarisability}) is then inconsistent.
Similarly, the factor
$q_\alpha(g) / P^{\rm sm}(g)$ appearing
in the KSB correction for PSF anisotropy
is, in general, dependent on the scale of the window used to
measure the PSF properties. 
\citeN{hfks98} have found from analysis of simulated images
that for the
HST WFPC2 instrument one can adjust the scale of the weight
function for the stars to render the calibration (\ref{eq:lkpolarisability}) 
and the KSB anisotropy correction reasonably
accurate, but 
it is not clear that this will apply in general.
Indeed, for diffraction limited seeing, the inadequacy of the
KSB formalism has a deeper root.  While the
assumption that the real PSF can be modeled as
a convolution of a circular PSF with some compact kernel $k(r)$ may be
a good approximation for atmospheric turbulence seeing with
small amplitude guiding errors and such-like, as we shall see,
this is not the case in general.

The current situation is therefore somewhat unsatisfactory.
In this paper we will develop an improved method of shear estimation
which does not suffer from the inadequacies noted above
and works for a PSF of essentially arbitrary form.
The layout of the paper is as follows:
In \S\ref{sec:operators} we construct an operator which
generalizes (\ref{eq:linearshearoperator}) to finite size PSF and which
generates the effect of a gravitational shear on the observed
(i.e.~post-seeing) surface brightness $\fobs$.
We first show that, quite generally, 
the effect on $\fobs = g \otimes \fint$ of a shear
applied to $\fint$ is equivalent to a shear applied directly to $\fobs$
plus a `commutator' term  which is a convolution
of $\fobs$ with a kernel $\gamma_\alpha H_\alpha(r)$ 
where $H_\alpha(r)$ may be computed from the PSF $g$.
We explore the properties of this kernel for various types and
PSF. 
\hide{ and find, for instance, that for seeing arising from
atmospheric turbulence the kernel is not compact, thus invalidating 
the calibration procedure using (\ref{eq:lkpolarisability}).
}
The kernel $H_\alpha$ involves $\ln \tilde g$ (where the tilde
denotes the Fourier transform), which appears, 
particularly in the case of
diffraction limited seeing, to be formally ill defined since $\tilde g$ vanishes
at finite radius.  We show however, that the operator which
generates the effect of a shear on a filtered
image which has had frequencies close to the diffraction limit
attenuated does not suffer from this problem.
In \S\ref{sec:estimators}
we specialize to 
weighted second moments, and compute their response to shear,
both for individual
objects \S\ref{subsec:individualresponse} and for the population of galaxies 
of a given flux, size and shape \S\ref{subsec:populationresponse}.
In \S\ref{sec:optimisation} we show how to optimally combine estimates of the 
shear from galaxies of various different types. The method is tested using
simulated mock images.
In \S\ref{sec:anisotropy} we discuss the correction for
PSF anisotropy,
and
propose a new technique involving measuring shapes from images which have
been convolved with a re-circularizing PSF.  
We draw attention to a hitherto ignored noise related bias
and show how this can be analyzed and corrected for.
\hide{In \S\ref{sec:acf} we show how these results can be
extended to calibrate shear estimators
derived from the auto-correlation function
of the sky brightness.}
In \S\ref{sec:discussion} we summarize the main results, and outline
how they can be applied to real data.
In the analysis here we will draw heavily on the
properties of real physical PSFs, and we include
as an appendix a brief review of basic PSF theory and discussion
of PSF properties as they relate to weak lensing observations.

\section{Finite Resolution Shear Operator}
\label{sec:operators}

We now consider how the shear operator (\ref{eq:linearshearoperator}) is modified
by finite resolution arising either in the atmosphere, telescope optics
or detector.  
Let the unperturbed surface brightness (i.e.~that which would have been
observed in the absence of lensing) be
\begineq
\fobs = g \otimes f
\endeq
so the perturbed surface brightness is
\begineq
\label{eq:perturbedsb}
\fobs' = g \otimes S_\gamma f
\endeq
Fourier transforming, and using the result that, at linear order
in $\gamma$, 
applying a shear in real space is
equivalent to applying a shear of the opposite sign in Fourier space
(i.e.~if $a = S_\gamma b$ then $\tilde a = S_{-\gamma} \tilde b$,
where $\tilde F (k)\equiv \int d^2 r \; F(r) e^{ik\cdot r}$),
we have
\begineq
\tilde \fobs' = \tilde g S_{-\gamma} \tilde f =
\tilde \fobs - \tilde g \delta S_{\gamma}(\tilde \fobs / \tilde g) 
\endeq
where  $\delta S_\gamma \equiv S_\gamma - 1$.
Since $\delta S_\gamma$ is a 1st order differential operator
we have $\delta S_\gamma(\tilde \fobs / \tilde g) = \tilde g^{-1} 
\delta S_\gamma \fobs - \tilde g^{-2} \tilde \fobs \delta S_\gamma \tilde g$,
and $\tilde g^{-1} \delta S_\gamma \tilde g = \delta S_\gamma \ln \tilde g$.
Consequently, to first order in $\gamma$ 
\begineq
\label{eq:deltatildefobs}
\tilde \fobs' = \tilde \fobs - \delta S_\gamma \tilde \fobs +
\tilde \fobs \delta S_\gamma \ln \tilde g  
\endeq
or, in real space,
\begineq
\fobs' = \fobs + \delta S_\gamma \fobs - (\delta S_\gamma h) \otimes \fobs
\endeq
where $h$ is the inverse  transform of
$\tilde h \equiv \ln \tilde g$, i.e.~the logarithm of the optical transfer function 
(OTF).
Invoking the definition of the shear operator 
(\ref{eq:linearshearoperator}), we have
\begineq
\label{eq:fobsoperator}
\fobs' = \fobs - \gamma_\alpha \Maij 
(r_i \d_j \fobs - (r_i \d_j h) \otimes \fobs).
\endeq
Note that the PSF is a real function so $\tilde g(-k) = \tilde g^*(k)$,
and this symmetry is shared by $\ln \tilde g$, so 
$h$ is also real.

The finite resolution shear operator (\ref{eq:fobsoperator}) 
is then $\fobs' = S_\gamma \fobs - (\delta S_\gamma h) \otimes \fobs$,
that is, it is the regular
shear operator $S_\gamma$ applied to the post-seeing image $\fobs$ plus
a `commutator term' $g\otimes S_\gamma f - S_\gamma(g \otimes f)$
which is a correction for finite PSF size and which is a
convolution of $\fobs$ with a kernel 
$\gamma_\alpha H_\alpha(r) = \gamma_\alpha \Maij r_i \d_j h = \delta S_\gamma h$ 
which one can compute
from the PSF $g$.  This seems quite promising; the effect of the
first term on the polarization of an object is just $\gamma_\alpha$ times the KSB
post-seeing shear polarizability.  The response of the polarization to the second term
should also be calculable; if the kernel is very compact then the
response will be given by the KSB linearized smear polarizability,
but even if it is not, 
(\ref{eq:fobsoperator}) should still allow one to compute the
polarization response since it expresses the change in $\fobs$,
and therefore in the polarization, or indeed any other statistic 
computable from $\fobs$,
directly in terms of the observed
surface brightness $\fobs$ itself.  
The finite resolution shear operator involves the function $h(r)$ which is the
transform of the logarithm of the OTF.  Since the OTF becomes exponentially
small or may vanish, two questions immediately arise:  Is the function $h(r)$
mathematically well defined? and can it be reliably computed from
PSFs measured from real stellar images? To address these questions we
now explore the form of
this `commutator kernel' $H_\alpha(r)$ for various types of PSF.

\subsection{Gaussian Ellipsoid PSF}
\label{subsec:gaussianshearoperator}

Consider first the simple though unrealistic
case of a Gaussian ellipsoid PSF: 
$g(r) = (2 \pi)^{-1} |m|^{-1/2} \exp(-r_i m_{ij}^{-1} r_j / 2)$,
where $m_{ij} = \langle r_i r_j \rangle$ is the matrix of
central second moments.
In this case, 
$\ln \tilde g(k) = -k_i m_{ij} k_j/ 2$, and so 
$\delta S_\gamma \ln \tilde g(k) = 
\gamma_\alpha M_{\alpha i m} m_{mj} k_i k_j$
and on transforming this we find that the kernel is the  
operator
$H_\alpha = M_{\alpha i m} m_{mj} \d_i \d_j$
(in which case the function $H_\alpha(r)$ can
be realized, for example, as the contraction of the
constant matrix $\gamma_\alpha M_{\alpha i m} m_{m j}$ with
the limit as $\sigma \rightarrow 0$ of the matrix of second partial
derivatives of a Gaussian ball $(2\pi \sigma^2)^{-1} \exp(- r^2 / 2
\sigma^2)$), and therefore
\begineq
\label{eq:GaussianPSFoperator}
\fobs' = \fobs - \gamma_\alpha \Maij 
(r_i \d_j \fobs + m_{il} \d_l \d_j \fobs).
\endeq
So, for a Gaussian PSF, the finite resolution shear operator is 
well defined and is a purely
local differential operator. 
Gaussian PSF's are, however unphysical, and do not
arise in real instruments.

\subsection{Atmospheric Turbulence}
\label{subsec:atmosphericshearoperator}

\begin{figure}[htbp!]
\centering\epsfig{file=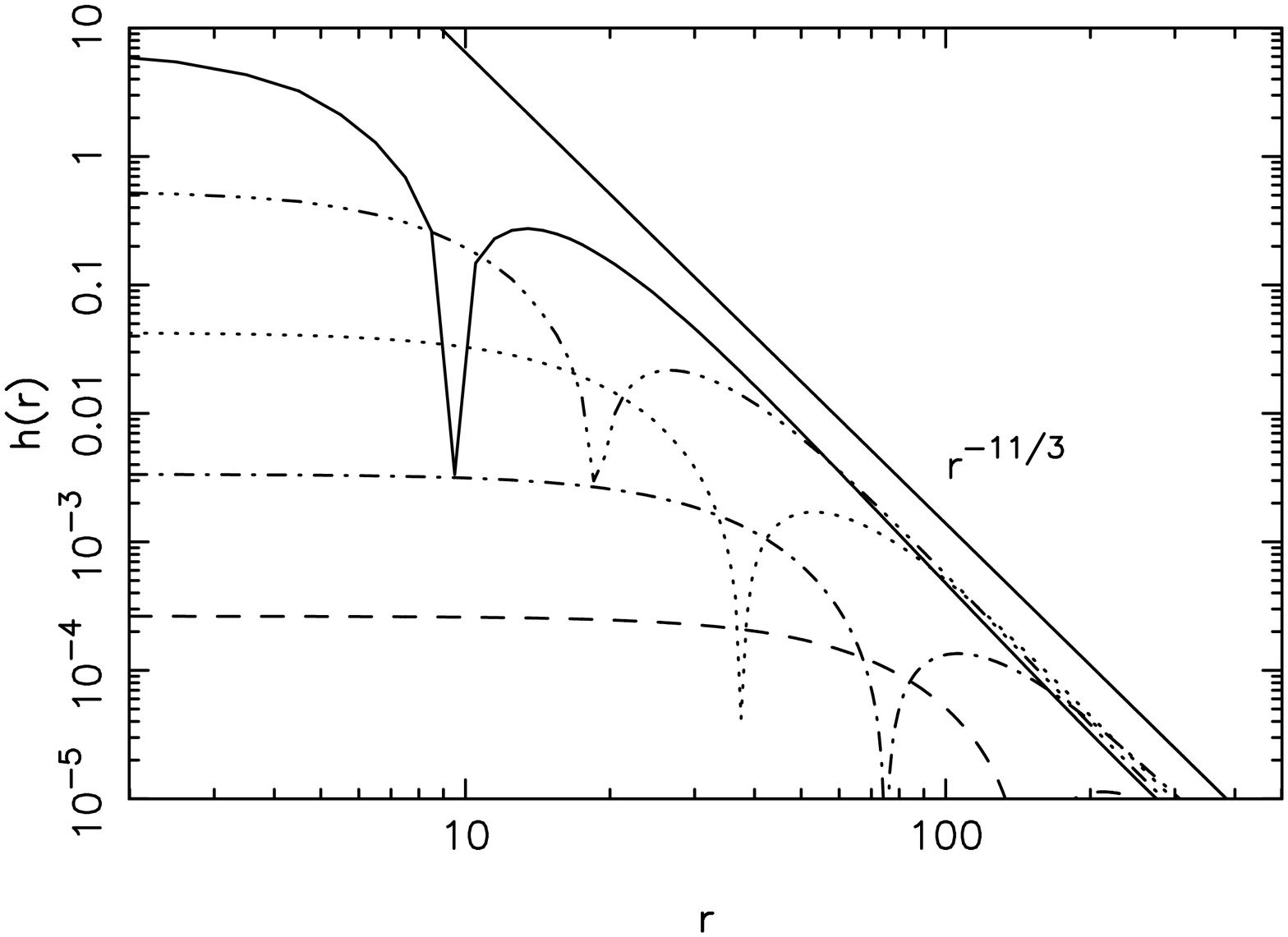,width={0.8 \linewidth},angle=0}
\figcap{Numerical calculation of 
$h(r; R) = \int d^2 k \; k^{5/3} \exp(-kR + i k \cdot r)$
for a range of $R$ values increasing by factors of 2. Also shown
is a pure $r^{-11/3}$ power law.
}
\label{fig:turbulenthkernel}
\end{figure}

Now consider atmospheric turbulence limited seeing. As reviewed in appendix \ref{sec:PSFtheory},
in that case 
$\tilde g(k) = \exp(- S(k D \lambda / 2 \pi)/2)$ where
for fully developed Kolmogorov turbulence the structure function is
$S(r) = 6.88 (r / r_0)^{5/3}$ with $r_0$ the Fried length, so in this case
$\tilde h \propto \ln \tilde g  \propto - k^{5/3}$. The commutator kernel
therefore involves the transform of this power-law, which diverges strongly
at high $k$, so how do we make sense of this?
 Dimensional 
analysis would suggest
that $h(r) = \int d^2k \; k^{5/3} e^{ik \cdot r}$
be a power law with
$h(r) \propto r^{-11/3}$.
The same argument, however, applied to a Gaussian
would suggest $h(r) \propto r^{-4}$, which we know to be
false as 
we have just shown that 
in that case $h(r)$ is just the second derivative of a 
$\delta$-function and has no extended tail.  
To clarify the situation, and to verify the validity of the
power-law form for $h(r)$ for atmospheric seeing,
consider the function $h(r;R)$ which is 
the transform  of
$k^{5/3}$ times an exponential cut-off function $\exp(-kR)$, that is
\begineq
h(r; R) \propto \int d^2 k k^{5/3} e^{-kR} e^{ik \cdot r}
\endeq
so $h(r)$ is the limit as $R \rightarrow 0$ of $h(r;R)$.
If we write this as an integral with respect to
rescaled wavenumber $y = kR$ it becomes clear that
$h(r;R)$ obeys a self-similar scaling with respect to choice of $R$ and
can be expressed in terms of some universal function $F(y)$ such
that $h(r;R) = R^{-11/3} F(r/R)$.  If we postulate that
$h(r;R)$ tends to some $R$-independent limit for finite
$r$ as $R \rightarrow 0$
then that limit must be a power law with $F(z) \propto z^{-11/3}$
so $h(r)$ must be proportional to $r^{-11/3}$.
This argument does not indicate the coefficient multiplying the power law which,
for a Gaussian, happens to vanish.
The value of $h$ at the origin is $h(0;R) = 2 \pi R^{-11/3} \Gamma(8/3)$,
i.e.~on the order of the value of the $r \gg R$ power law asymptote extrapolated to
$r \sim R$ (note that for the Gaussian the analogous gamma function
is not defined).
A numerical integration for various values of $R$ is shown
in figure \ref{fig:turbulenthkernel}.  This shows that as
we decrease $R$, the function $h(r;R)$ does indeed tend to
a $r^{-11/3}$ power law with finite non-zero $R$-independent amplitude,
but that the power law breaks (with a change of sign) at $r \sim R$
and becomes asymptotically flat for $r \ll R$.
For finite $R$ this 
function is well characterized as a positive `softened $\delta$-function'
core with width $\sim R$, central value $\sim R^{-11/3}$ and therefore with
weight proportional to $R^{-5/3}$, surrounded by a negative power
law halo with $h \propto r^{-11/3}$.  In the limit $R \rightarrow 0$, the
core shrinks and becomes negligible, leaving only the power law halo.
This power law has the same slope as 
the well known large angle
$g(r) \propto r^{-11/3}$ form
of the PSF, but the PSF departs from this law at angular scale $r_g \sim \lambda / r_0$ corresponding
to the Fried length, whereas $h(r)$ is a 
perfect power law and shows no features at the Fried scale.

The function $h(r)$ diverges strongly at the origin and the same is
true of the kernel $H_\alpha = M_{\alpha i j} r_i \d_j h$, so
$H_\alpha = - 11/3 r^{-11/3} \{\cos 2 \varphi, \sin 2 \varphi\}$.
This does not give rise to any inconsistency with
(\ref{eq:fobsoperator}) however. The observed sky $\fobs$ is coherent on the scale
of the PSF $r_g$, so in computing the contribution to the commutator term
\begineq
\delta \fobs = \gamma_\alpha M_{\alpha i j} \int d^2r' \; r'_i r'_j (r')^{-17/3}
\fobs(r - r')
\endeq 
at small $r' \ll r_0$, 
we can perform a Taylor expansion of $\fobs$, and we find that the
first non-vanishing term is the second order term
$\delta \fobs \sim (\d_i \d_j \fobs)_r
\int d^2r'\; r'_i r'_j (r')^{-11/3}$ which has no physical divergence at $r' = 0$.
Somewhat unfortunately perhaps, the same line of argument shows that the commutator
term cannot in this case be assumed to be a convolution with
some compact function $k(r)$.
A necessary condition for this to be valid is that
the unweighted second moment of the kernel 
$p_\alpha = \Malm \gamma_\beta \int d^2r\; r_l r_m H_\beta(r)$ should
be well defined and tend to a finite limit within some
small radius $\ll r_g$, the characteristic width of the PSF.  
But this is the same integral as above, which
does not tend to any well defined value, but rather diverges as the
$1/3$ power of the upper limit on the integration radius.
Thus while the $h(r) \propto r^{-11/3}$ asymptote
is quite steep, it is not sufficiently steep to render the
$p_\alpha$ value well defined. 
This strictly invalidates the
approximation of \citeN{lk97} for turbulence limited seeing.

The Kolmogorov law is only expected to apply over a finite range
of scales. The structure function will fall below the
$r^{5/3}$ form at the `outer scale', which recent measurements 
at La Silla suggest to
be typically on the order of 20m (\citeN{mtz+98}, though see also the review
of earlier results in \citeN{azb+97}). 
Fast guiding (often referred to as `tip-tilt' correction) would also
effectively reduce the structure function
at these frequencies, and such effects act in a similar
manner to the simple exponential cut-off we have assumed.  To compute the
OFT properly, one should include the effect of the aperture.
The detailed form
of $h(r)$ at very small $r$ will be sensitive to the detailed form of the
outer scale cut-off and/or aperture, but
provided these lie at scales much greater than the
Fried length (which is the case for large aperture telescopes at good sites)
the effect on the commutator term should be almost independent of the
cut-off because any signal at the relevant spatial frequencies
will have been attenuated by a large factor.
There will also be deviations from Kolmogorov structure function
law at small separation; 
diffusion will damp out fine scale turbulence, and mirror roughness
will add additional high spatial frequency phase errors.
These effects will modify the $h \propto r^{-11/3}$ halo
at large $r$, just as they modify the large-angle $g  \propto r^{-11/3}$
form of the PSF.   These effects may profoundly influence the
behavior of $h$, $H_\alpha$ at  large angle,  but
have little impact on the type of shape statistics considered here,
where the large-angle contribution to the polarization is suppressed
by the weight function or the isophotal cut-off. 

\subsection{Diffraction Limited Seeing}
\label{subsec:diffractionshearoperator}

Let us now consider 
diffraction limited observations. 
In this case the OTF $\tilde g$ 
falls to zero at finite spatial frequency --- the diffraction
limit --- and so $\ln \tilde g$ diverges as one approaches the diffraction
limit from below and is not defined
for higher frequencies.
\hide{This divergence is similar to that which afflicts the deconvolution
operator $\tilde g^{-1}$, and immediately raises worries about
whether convolution with $H$ would be well defined and the
possibility of amplification of noise.}
It is easy to understand how this divergence arises physically.
As noted above, applying a weak shear in real space is equivalent 
to applying a shear of the
opposite sign in Fourier space. Thus the information in some Fourier
mode of the sheared image comes from a slightly displaced mode
in the unsheared image. If the OTF $\tilde g$
is finite and continuous,
as is the case for atmospheric turbulence limited
seeing, then the image $\fobs$ contains all of the information required
to predict $\fobs'$.  For diffraction limited seeing however,
and for observations through a narrow band filter,
the OTF has a well defined edge, so the information contained in $\fobs$ for
spatial frequencies just inside the cut-off may lie outside the cut-off
in $\fobs'$ and so will be missing. For a finite band pass the form of the OTF
will be modified, but must still fall to zero at the diffraction limit for the
highest frequencies passed by the 
filter and $\ln \tilde g$ is still formally divergent.

Thus, in general, from knowledge of $\fobs = g \otimes f$, it is 
strictly speaking not possible to say how $\fobs$ would change in
response to a small but finite shear applied before seeing; 
there are Fourier modes within a distance $\delta k \sim \gamma k_\max$ of
the diffraction limit that one cannot predict.  As an extreme example,
imagine we have a pure sinusoidal ripple on the sky which lies just outside
the diffraction limit and which is therefore invisible. 
Applying an appropriate shear can bring that mode inside the
limit and the ripple will appear as if from nowhere.
As applied to real signals, however, 
this formally divergent behavior does not
present a serious problem.  First, if we observe
galaxies of overall extent $r_G$, which is typically not much larger than the
seeing disk, then the transform must vary smoothly with
coherence scale $\delta k \sim 1/ r_G$
which, for sufficiently small $\gamma$ will greatly exceed
$\gamma k_\max$; this rules out the possibility of
isolated spikes lurking just beyond the diffraction limit as in the example. 
Also, for filled aperture optical telescopes
the marginal modes are quite 
strongly attenuated, and, for any finite measurement
error from e.g.~photon counting statistics, will contain very little
information.  The situation here is similar to that encountered
in analyzing atmospheric turbulence PSF; there, while the very small scale details
of $h(r)$ are sensitive to the aperture or outer scale cut-off, when applied to
real data they have essentially no effect.  Similarly, if one can generate
a function which accurately coincides with $\tilde h = \log \tilde g$ where 
$g(k)$ exceeds some small value, but which tapers off smoothly at larger
frequencies (rather than diverging at $k_\max$), then this should
have a well defined transform and should give what is, for all practical
purposes, a good approximation to the true finite resolution shear
operator. 

One way to explicitly remove the divergence is to compute
shapes from an image which has had the marginally detectable modes
attenuated.
If one re-convolves the observed field $\fobs$ with some filter function
$\gdag(r)$ to make 
a smoothed image $\fs = \gdag \otimes \fobs$ then in Fourier space we have
\begineq
\tilde \fs' = \tilde \fs - \tilde \gdag \delta S_\gamma \tilde \fobs +
\tilde \fobs \tilde \gdag \tilde g^{-1} \delta S_\gamma \tilde g  
\endeq
so provided $\tilde \gdag$ falls off at least as fast as $\tilde g$ as one approaches the
diffraction limit this operator
is well defined. In real space the corresponding operator is 
\begineq
\label{eq:fsoperator0}
\fs' = \fs - \gamma_\alpha \Maij (\gdag \otimes ( r_i \d_j \fobs) 
- \gdag \otimes (r_i \d_j h) \otimes \fobs)
\endeq
\hide{
\begineq
\fs' = \fs - \gamma_\alpha (\gdag \otimes (\Maij r_i \d_j \fobs)
- H_\alpha \otimes \fobs)
\endeq
}
so now the commutator term is the
convolution of $\fobs$ with $ H_\alpha = \Maij \gdag \otimes (r_i \d_j h)$.
In principle one can design $\gdag$ such that $\fs$ is essentially
identical to $\fobs$; simply set $\tilde \gdag$ to unity for all modes
within some tiny distance of the diffraction limit
and zero otherwise.  However,
this may not be a very good idea; the sharp edge of
such a filter function in $k$-space will result in 
ringing in real space, both in the filter $\gdag(r)$ and especially in $H_\alpha(r)$.
These extended wings have no effect on real signal, but will
couple to the incoherent noise in the images, whose spectrum does
not  fall
to zero as one approaches
the diffraction limit.  
\hide{
There are two possible resolutions to this
obstacle. First, rather than trying to compute the polarizability
of every object individually, one can average the polarization
of galaxies binned by by flux, size and eccentricity say, and
also compute the polarizability for the mean profile, thus
effectively avoiding problems with correlated noise.
}
To ameliorate these problems one
can choose $\tilde \gdag$ to have a soft roll-off as one 
approaches $k_\max$ to counteract the divergence
of $1/\tilde g$.  
\hide{The optimal choice of filter $\tilde \gdag$ must
necessarily depend on the type of signal one is dealing with,
but we suspect that in most cases it corresponds to a smoothing
quite similar to the PSF itself.  }
One simple option
is to set $\gdag = g$; i.e.~to re-convolve with the
PSF itself. In this case $H_\alpha(r)$ will be about 
as compact as the PSF,
and we then have
\begineq
\label{eq:fsoperator1}
\fs' = \fs - \gamma_\alpha \Maij (g \otimes (r_i \d_j \fobs) -
(\fobs \otimes (r_i \d_j g))
\endeq
or equivalently
\begineq
\label{eq:fsoperator2}
\fs' = \fs + \gamma_\alpha \Maij 
(2 (r_i \d_j g) \otimes \fobs -  r_i (\d_j g \otimes \fobs))
\endeq
where we have integrated by parts to avoid explicitly
differentiating the image $\fobs$.  

To summarize, we have shown in (\ref{eq:fobsoperator}) that
the effect on the observed sky $\fobs$ 
of a weak shear applied before seeing
can be written as a shear applied after seeing plus a convolution
with some kernel which 
is the sheared transform of the logarithm of the OTF, which
can be computed from the
PSF. We have explored the form of this kernel for both Gaussian and
more realistic models for the PSF.  For atmospheric turbulence limited
seeing the kernel is a power law $H \sim r^{-11/3}$.  
For diffraction limited seeing the shear operator appears formally ill-defined
but this is not a serious problem when applied to
real data and that one can, for example, compute the operator for a slightly
smoothed sky $f_s = \gdag \otimes \fobs$ in a divergence free manner.
We have presented the shear operator for the case where $\gdag = g$.
This could potentially be used to compute the response of
the shape statistics measured by the FOCAS and/or Sextractor
packages, as these are measured from just such a re-convolved image, 
but we will not pursue this here.

\section{Weighted Moment Shear Estimators}
\label{sec:estimators}

We now specialize to weighted quadrupole moments as defined in
(\ref{eq:weightedmomentdefinition})
and compute how these respond to shear.
We first compute the response of the moments of an individual
object, and we then compute the conditional mean response for a population
of objects having given flux, size etc.
This will allow us to compute an optimal weight function for
combining shear estimates from galaxies of a different fluxes, sizes
and shapes.  

\subsection{Response of Weighted Moments for Individual Objects}
\label{subsec:individualresponse}

Consider again for illustration the case of a Gaussian ellipsoid PSF
$g(r) \propto \exp(-r_i m^{-1}_{ij} r_j/2)$ and moments 
$q_{lm} = \int d^2r \; \fobs(r) w(r) r_l r_m$.  From
(\ref{eq:GaussianPSFoperator}), (\ref{eq:weightedmomentdefinition})
we have $q'_{lm} = q_{lm} + \delta q_{lm}$ with
\begineq
\delta q_{lm} = - \gamma_\alpha \Maij 
\int d^2r\; w r_l r_m (r_i \d_j \fobs + m_{ip} \d_p \d_j \fobs)
\endeq
Integrating by parts to replace derivatives of $\fobs$ with
derivatives of $wr_l r_m$   
we find the linear response of $q_A \equiv {1 \over 2} M_{Alm} q_{lm}$
to a shear can be written
\begineq
\label{eq:weighteddeltap}
\delta q_A = P_{A\beta} \gamma_\beta
\endeq
with `shear polarizability'
\begineq
\label{eq:polarisability}
P_{A\beta}  = \int d^2 r \; \P_{A\beta}(r) \fobs(r)
\endeq
and where
\begineq
\label{eq:Pdefinition}
\P_{A\beta}(r) = {1\over 2} M_{Alm} M_{\beta ij} [r_i \d_j w r_l r_m - m_{ip} \d_p \d_j 
w r_l r_m] 
\endeq
For the special case of $w = 1$, i.e.~unweighted moments, we find 
$\P_{\alpha\beta}
= \delta_{\alpha\beta}(r_i r_i - m_{ii})$
and hence $\delta q_\alpha = \gamma_\alpha (q_{ii} - m_{ii})
= 2 \gamma_\alpha (q_0 - m_0)$
in accord with the result 
obtained in the Introduction.  In this case, $P_{\alpha\beta}$
is a combination of zeroth and second moments of $\fobs$.

\begin{figure}[htbp!]
\centering\epsfig{file=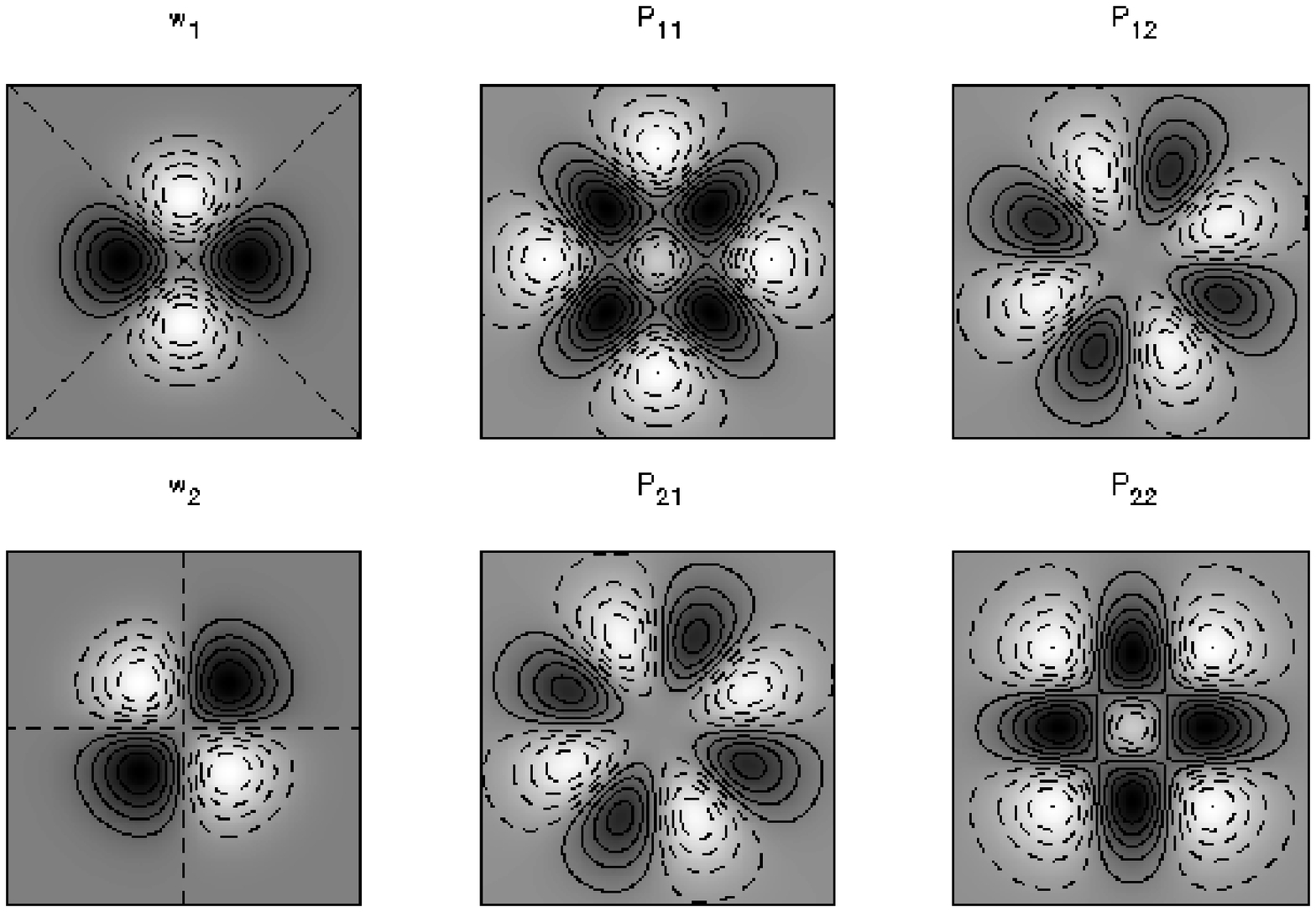,width={1.0 \linewidth},angle=0}
\figcap{The panels on the left show the pair of
functions $w_1$, $w_2$ with
$w_\alpha(r) = {1 \over 2} \Malm w(r) r_l r_m$ which
when multiplied  by $f_s$ and integrated give the polarization statistic
$q_\alpha$.
The four panels on the right show the components of the polarizability
kernel $\P_{\alpha\beta}(r)$ which when multiplied by $\fobs$ and
integrated yields the polarizability $P_{\alpha\beta}$.  
The PSF was computed from a turbulence limited
OTF $\tilde g = \exp(-0.5(kr_*)^{5/3})$, and the smoothing kernel
was $w(r) = \exp(-0.5(r/r_*)^{2})$ with scale length $r_*$ equal to
one eighth of the box side.
}
\label{fig:PSFkernel}
\end{figure}

For a general PSF and for moments measured from a filtered field 
$f_s = \gdag \otimes \fobs$ as
a weighted moment
\begineq
\label{eq:polarisationfromfs}
q_A = {1\over 2} M_{Alm} \int d^2r \; w(r) r_l r_m f_s(r)
\endeq
we find
using (\ref{eq:fobsoperator}) that the response $\delta q_A$ can be cast in the same
form, but now with
\begineq
\label{eq:Pab1}
\P_{A \beta}(r) ={1\over 2} M_{Alm} M_{\beta ij} 
[r_i ( \gdag \oplus (\d_j \omega r_l r_m))
- (r_i h) \oplus \gdag \oplus (\d_j \omega r_l r_m)]
\endeq
where we have defined the correlation operator
$\oplus$ such that $(a \oplus b)_r \equiv \int d^2 r'
a(r') b(r' + r)$.
If the moments are measured directly from the unfiltered images $\fobs$ then one
can replace $\gdag(r)$ with a Dirac $\delta$-function to obtain
\begineq
\label{eq:Pab2}
\P_{A \beta}(r) ={1\over 2} M_{Alm} M_{\beta ij} [r_i \d_j w r_l r_m  + (hr_i) \oplus
(\d_j w r_l r_m)]
\endeq
whereas for the special case $\gdag = g$
\begineq
\label{eq:Pab3}
\P_{A \beta}(r) = {1 \over 2} M_{Alm} M_{\beta ij} 
[2 r_i (g \oplus \d_j(w r_l r_m)) - g \oplus
(r_i \d_j(w r_l r_m)) ].
\endeq
The function $\P_{\alpha\beta}(r)$ 
is shown in figure \ref{fig:PSFkernel}
for a turbulence limited PSF and for a Gaussian window function $w(r)$.
Note that (\ref{eq:Pab1}) and (\ref{eq:Pab3}) are well defined continuous functions even in the
limit that the weight function becomes arbitrarily small; i.e.~$w(r) \rightarrow \delta(r)$.

Equation (\ref{eq:weighteddeltap}), along with the appropriate expression
for $P_{A \beta}(r)$
tells us how the polarization statistic
for an individual object 
formed from weighted quadrupole moments responds to
a gravitational shear.  This is an essential ingredient in
calibrating the shear-polarization relation for a population
of galaxies.  As we shall see in the next section, there are some
subtleties involved, but for now we note that if we
simply average (\ref{eq:weighteddeltap}) over all galaxies on
a patch of sky we have
\begineq
\langle q'_\alpha \rangle - \langle q_\alpha \rangle = 
\langle P_{\alpha \beta} \rangle  \gamma_\beta
\endeq
so an estimate of the net shear is given by
\begineq
\label{eq:gammaestimator}
\hat \gamma_\alpha =  
\langle P_{\alpha \beta} \rangle^{-1} (\langle q'_\beta \rangle - \langle q_\beta \rangle)
\endeq
Which is the generalization of (\ref{eq:calibratedsimpleestimator})
to weighted moments.  
The term $\langle q'_\beta \rangle$ 
in (\ref{eq:gammaestimator}) is the
averaged observed polarization.  The term $\langle q_\alpha \rangle$
is the mean polarization generated by anisotropy
of the PSF, and we will show how this can be dealt with below.

Ideally, in averaging polarizations, one should
apply weight proportional to the square of the signal to
noise ratio, which one would expect to be a function of the flux,
size, eccentricity etc.~of the objects.  If the shape is measured 
with some fairly compact window function $w(r)$,
then the total flux, which may be dominated by the profile of the object at
considerably larger radius, is probably not ideal and one will 
likely obtain better
performance if one takes the weight function to be a function of
$q_0$, $q^2 \equiv q_\alpha q_\alpha$ and a weighted flux
\begineq
F = \int d^2r \; w(r) f_s(r).
\endeq
The response of $F$ can be computed in much the same
way as $q_A$, and we find from (\ref{eq:fsoperator0})
$\delta F = R_\alpha \gamma_\alpha$ with $R_\alpha = 
\int d^2 r \; \R_\alpha \fobs(r)$ with
\begineq
\R_\alpha(r) = M_{\alpha i j}
[r_i (\gdag \oplus \d_j w) - (r_i h) \oplus \gdag \oplus \d_j w)]
\endeq
or, for the case $\gdag = g$,
\begineq
\R_\alpha(r) = M_{\alpha i j}
[2 r_i (g \oplus \d_j w) - g \oplus \d_j w r_i)].
\endeq

In the foregoing we have implicitly assumed that applying
a shear does not affect the location of an object.  This is
not necessarily the case.  If objects are detected as peaks of
the surface brightness $\fobs$ smoothed with some detection
filter $w_d$, that is as peaks of $f_d = w_d \otimes \fobs$, then
for an object which in the absence of shear lies at the origin,
we have $d_i(0) = \d_i (w_d \otimes \fobs)_0 = 
\int d^2 r \; \d_i w_d(r) \fobs(r) = 0$ while after applying
a shear we have $d_i(0) = \gamma_\alpha \Malm
\int d^2 r \; \d_i w_d [r_l \d_m \fobs - (r_l \d_m h) \otimes \fobs]$.
This will not, in general, vanish, implying that the peak location will
have shifted, and consequently the central second moments should
be measured about the shifted peak location, whereas in the above
formulae we have computed the change in the moments without
taking the shift into account.  One could incorporate this effect,
but at the expense of considerable complication of the results.
There is some reason to think that this effect is rather weak.
In particular, if the galaxy is symmetric under rotation by
180 degrees, so $\fobs(r) = \fobs(-r)$, then the shift in the
centroid vanishes.  In general this is not the case, and 
the formulae above should be considered only an approximation.

\subsection{Response of the Population}
\label{subsec:populationresponse}

Equation (\ref{eq:gammaestimator}) above gives a properly calibrated estimate of the
gravitational shear. It is, however, less than ideal as
the polarization average is taken over all galaxies with equal weight.  
This is neither desirable
nor is it achievable in practice due to selection limits, 
and what one would rather have is an expression for the
average induced polarization for all galaxies in some cell of
flux, size and shape space, which we parameterize by
$F$, $q_0$, and $q^2 \equiv q_\alpha q_\alpha$. One can
then average appropriately weighted combinations of the
average shear estimate for each cell.  The mean induced
polarization for such a cell depends not only on the polarizabilities
of the objects contained therein, but also on the gradient of the
mean density of objects as a function of 
the photometric parameters $F$, $q_0$, $q_\alpha$. Consider
a slice through this 4-space at constant $F$, $q_0$.  A shear will
induce a general flow of particles in this space in
the direction $\hat q_\alpha = \hat \gamma_\alpha$. 
The mean polarisation for a cell in $F-q_0-q^2$
is the average around an annulus in $q_\alpha$ space, and
depends quite sensitively on the local slope of the
distribution of particles in $|q_\alpha$.
These factors can have a profound influence on the
weighting scheme; for a distribution which is flat near the
origin in $q_\alpha$ space, like a Gaussian for example, nearly circular
objects have no response and should therefore receive no weight.
For a randomly oriented distribution of circular disk galaxies,
in contrast, the distribution in  $q_\alpha$-space has a cusp
at the origin; the response becomes asymptotically infinite, and
these objects dominate the optimally weighted combination.
These examples are both idealized, but underline the importance of
computing the population response in order to obtain
optimal signal to noise.

Let us first compute the conditional mean polarization for
galaxies of a given flux and size:
$\langle q_\alpha \rangle_{F,q_0}$. 
The mapping of the photometric parameters $F$, $q_0$, $q_\alpha$ is
\begineq
\label{eq:Fq0qmapping}
\begin{matrix}{
F' = F + R_\beta \gamma_\beta\cr
q'_0 = q_0 + P_{0\beta} \gamma_\beta \cr
q'_\alpha = q_\alpha +  P_{\alpha \beta} \gamma_\beta 
}\end{matrix}
\endeq
so we need to consider the distribution of galaxies in 
$F,\;q_0,\;q_\alpha,\;R_\alpha,\;P_{0\alpha},\;P_{\alpha \beta}$ with
lensed and unlensed distribution functions related by
\begineq
n'(F', q'_0, q'_\alpha, R'_\alpha, P'_{0\alpha}, P'_{\alpha \beta})
dF' dq'_0 d^2 q' d^2 R d^2 P'_0 d^4 P' =
n(F, q_0, q_\alpha, R_\alpha , P_{0\alpha}, P_{\alpha \beta})
dF dq_0 d^2 q d^2 R d^2 P_0 d^4 P.  
\endeq
Multiplying by $W(F', q'_0) q'_\alpha$, where 
$W$ is some arbitrary 
function, 
and integrating over all variables we have
\begineq
\int dF' dq'_0 d^2 q' d^2 R d^2 P'_0 d^4 P' n' W(F', q'_0) q'_\alpha
= \int dF dq_0 d^2 q d^2 R d^2 P_0 d^4 P n
W(F + \delta F, q_0 + \delta q_0) (q_\alpha + \delta q_\alpha). 
\endeq
Now to zeroth order in $\gamma$ this vanishes because of statistical
anisotropy of the unlensed population, so using (\ref{eq:Fq0qmapping}) for $\delta F$ etc.~and
performing a Taylor expansion
of $W(F + \delta F, q_0 + \delta q_0)$ and integrating by parts we have
\begineq
\int dF' 
\hide{dq'_0 d^2 q' d^2 R' d^2 P'_0}
\ldots  d^4 P' \; n' W(F', q'_0) q'_\alpha
= \gamma_\beta \int dF 
\hide{dq_0 d^2 q d^2 R d^2 P_0}
\ldots  d^4 P \;
W(q_0) \left(n P_{\alpha \beta} - P_{0\beta} q_\alpha {\d n \over \d q_0}
- R_\beta q_\alpha {\d n \over \d F} \right).
\endeq
To first order in $\gamma$  we can replace unprimed by primed
quantities throughout the integral on the RHS since we need to
compute this only to zeroth order accuracy.  We now have a relation between the
mean value of the observed polarization on the LHS
and some other observable, again
integrated over the distribution of observed galaxy properties (rather than of the unlensed parent
distribution).
Since $W(q_0)$ is arbitrary, and dropping primes, this implies
\begineq
\int d^2 q d^2 R d^2 P_0 d^4 P n q_\alpha
= \gamma_\beta \int \cdots \int d^2 q d^2 R d^2 P_0 d^4 P 
\left(n P_{\alpha \beta} - P_{0\beta} q_\alpha {\d n \over \d q_0}
- R_\beta q_\alpha {\d n \over \d F} \right)
\endeq
or equivalently, that the conditional average polarization is
\begineq
\langle q_\alpha \rangle_{F, q_0} = \gamma_\beta
\left[\langle P_{\alpha \beta} \rangle - {1 \over n} {\d n \langle P_{0\beta} q_\alpha \rangle \over \d q_0}
- {1 \over n} {\d n \langle R_{\beta} q_\alpha \rangle \over \d F}
\right]_{F, q_0}
\endeq
where $n = n(F, q_0)$.
Thus, as expected, the shear induced shift in the mean polarization for 
galaxies of a given size and flux 
differs from the mean of the shift $\delta q_\alpha = P_{\alpha \beta} \gamma_\beta$ for
an individual object.  This is because a shear
changes the weighted flux and size of an object in a way which is correlated with its
shape.  When we average the
shear for galaxies in some cell in flux-size space we are averaging
over galaxies which have been scattered in size and flux and we obtain
a bias in the net polarization which depends on the gradients of the
distribution function.


We can generalize this to compute the response for galaxies 
of given flux $F$, size $q_0$, and rotationally invariant shape parameter
$q^2 = q_\alpha q_\alpha$.  To do this, we set 
$q_\alpha = q \hat q_\alpha$ with the unit polarization vector 
$\hat q_\alpha = \{\cos \varphi, \sin \varphi\}$, so
$d^2q = q dq d\varphi$. 
We then have, now for some arbitrary function $W(F, q_0, q^2)$,
\begineq
\int dF' dq'_0 d{q^2}' d\varphi' d^2 R d^2 P'_0 d^4 P' n' 
W(F', q'_0, {q^2}') q'_\alpha
= \int dF dq_0 dq^2 d\varphi d^2 R d^2 P_0 d^4 P n
W(F + \delta F, q_0 + \delta q_0, q^2 + \delta q^2) 
(q_\alpha + \delta q_\alpha) 
\endeq
where now $n = n(F, q_0, q^2, \varphi, R_\alpha, P_{0\alpha}, P_{\alpha \beta})$.
Using $\delta F$ etc.~from (\ref{eq:Fq0qmapping}) and $\delta q^2 = 2 q_\eta P_{\eta \beta} \gamma_\beta$ 
we have
\begineq
\int d\varphi d^2 R d^2 P_0 d^4 P n q_\alpha
= \gamma_\beta \int d\varphi  d^2 R d^2 P_0 d^4 P 
(n P_{\alpha \beta} -R_\beta q_\alpha \d n / \d F - P_{0\beta} q_\alpha \d n / \d q_0
- 2 P_{\eta \beta}  \d n q_\eta q_\alpha / \d q^2)
\endeq
or equivalently $\langle q_\alpha \rangle_{F, q_0, q^2} = \Peff_{\alpha \beta} \gamma_\beta$
with effective polarizability
\begineq
\label{eq:Peff1}
\Peff_{\alpha \beta} = \langle P_{\alpha \beta} \rangle 
- {2 \over n} {\d n \langle q_\eta P_{\eta \beta} q_\alpha \rangle \over \d q^2}
- {1 \over n} {\d n \langle P_{0\beta} q_\alpha \rangle \over \d q_0}
- {1 \over n} {\d n \langle R_{\beta} q_\alpha \rangle \over \d F}
\endeq
where now $n = n(F, q_0, q^2)$, and all averages are at fixed $F, \; q_0, \; q^2$.

The photometric parameters $q_0, q_\alpha$ appearing here are 
unnormalized.  This is convenient for computing the linear response
functions.  It is, however, somewhat awkward here since the distribution
function $n(F, q_0, q^2)$ is highly skewed since $q_0$ and $q^2$ correlate
very strongly with the flux.  In computing the effective polarizability
it is more convenient to work with rescaled variables $q_0' = q_0 / F$, and
${q^2}' = q^2 / F^2$.  The distribution function in rescaled variables is
\begineq
n'(F, q_0', {q^2}') = F^3 n(F, q_0, q^2).
\endeq
If we also re-scale the polarizabilities $R_\alpha' = R_\alpha / F$,
$P_{A\beta}' = P_{A\beta} / F$, $\Peff_{A\beta}' = \Peff_{A\beta} / F$,
re-express (\ref{eq:Peff1}) entirely in terms of primed quantities and
then drop the primes
we find
\begineq
\label{eq:Peff2}
\Peff_{\alpha \beta} = \langle P_{\alpha \beta} \rangle 
- {2 \over n} {\d n \langle q_\eta P_{\eta \beta} q_\alpha \rangle \over \d q^2}
- {1 \over n} {\d n \langle P_{0\beta} q_\alpha \rangle \over \d q_0}
+ {1 \over n} \left[1 - {\d \over \d \ln F} + {\d \over \d \ln q_0} + 2 {\d \over \d \ln q^2}
\right] n \langle R_\beta q_\alpha \rangle
\endeq
where we have used the result that for any function
$X(F, q_0', {q^2}')$ the partial derivative WRT $F$ at constant $q_0$, $q^2$ is
$(\d X(F, q_0', {q^2}') / \d F)_{q_0,q^2} = \d X / \d F - (q_0'/F) \d X / \d q_0' -
(2 {q^2}'/F) \d X /\d {q^2}'$ to compute the term involving $R_\alpha$.
The rather cumbersome expression (\ref{eq:Peff2}) calibrates 
the relation between the shear and the
mean polarization for galaxies in a small cell in flux-size-shape
space.  To compute it we need to bin galaxies in this space to
obtain the mean density $n$ and the various averages appearing here,
and then perform the indicated partial differentiation.  The form 
(\ref{eq:Peff2}) is somewhat inconvenient as the density $n(F, q_0, q^2)$
is asymptotically constant as $q^2 \rightarrow 0$ and one has to
properly deal with the discontinuous
derivative at this boundary. A computationally
more convenient approach is to make one final transformation from
$q^2 \rightarrow q$; since $n(F, q_0, q) = 2 q n(F, q_0, q^2)$ falls to
zero as $q \rightarrow 0$, and there is then no need for any special treatment of the
derivatives at the boundary.  With this transformation we have
\begineq
\label{eq:Peff3}
\Peff_{\alpha \beta} = \langle P_{\alpha \beta} \rangle 
- {1 \over n} {\d n \langle q_\eta P_{\eta \beta} \hat q_\alpha \rangle \over \d q}
- {1 \over n} {\d n \langle P_{0\beta} q_\alpha \rangle \over \d q_0}
+ {q \over n} \left[1 - {\d \over \d \ln F} + {\d \over \d \ln q_0} +  {\d \over \d \ln q}
\right] n \langle R_\beta \hat q_\alpha \rangle
\endeq
where now $n = n(F, q_0, q)$.


What about measurement noise?  
Let us assume that one has been given an image containing signal and
measurement noise, and that one has detected objects, and measured
quantities like $F$, $q_A$.  How would these photometric
parameters change under the influence of a gravitational shear?
The answer is given by (\ref{eq:Fq0qmapping}), but with the
understanding that $R_\beta$, $P_{0\beta}$ etc.~be the
response functions one would measure in the absence of
noise. This means that (\ref{eq:Peff3}) is also applicable
with the same proviso. The major terms in (\ref{eq:Peff3}) are
however invariant of additive noise. The exceptions are
the terms invoving  $\langle P_{0\beta} q_\alpha \rangle$ and
$\langle R_\beta q_\alpha \rangle$ which are quadratic
in the sky surface brightness. The measured expectation
values $\langle P_{0\beta} q_\alpha \rangle$ etc.~therefore 
exceed the true values, but by an amount one can
calculate from the known properties of the measurement
noise. 
Another implicit assumption in the above analysis is that
the objects are actually detected, which restricts
applicability to objects which are detected at a reasonable
level of significance.  Aside from this, the results
above should be applicable in the presence of measurement
noise.
To test these claims we have made extensive
simulations with mock data, the details of which are described in
appendix \ref{sec:simulation}.
Figure \ref{fig:Pplot} shows the results of one of these. The actual
polarization agrees quite closely with the effective polarizability,
even for very faint objects.  While the differences between the
effective polarizability and that for individual objects is
not very large - typically on the order of 20\% or so - the
effective polarizability clearly describes the true response
more faithfully.
We shall now use (\ref{eq:Peff3}) to construct a minimum variance weighting scheme
for combining shear estimates. 

\begin{figure}[htbp!]
\centering\epsfig{file=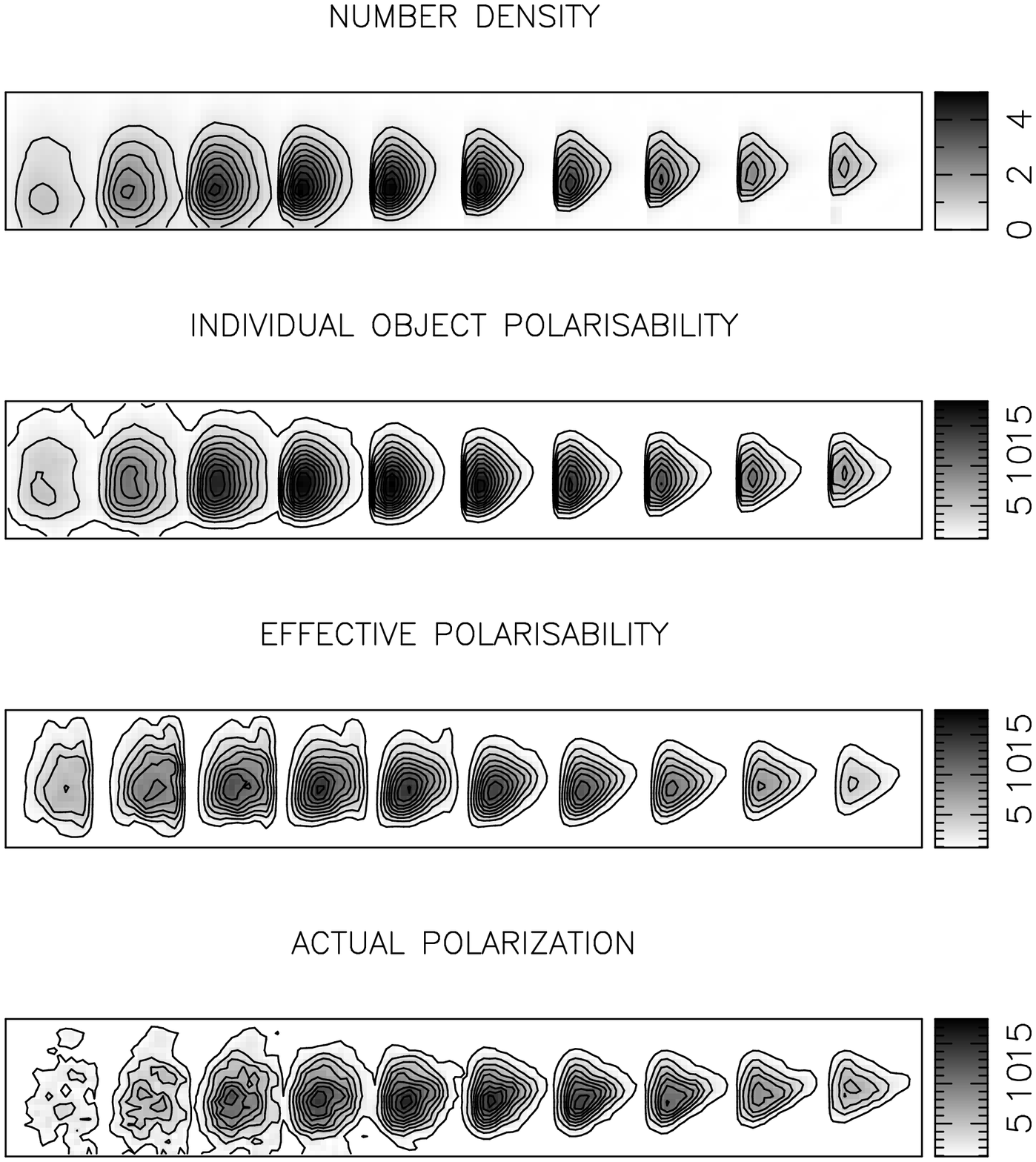,width={1.0 \linewidth},angle=0}
\figcap{The top panel shows the density of galaxies detected in the simulations
described in appendix \ref{sec:simulation}
as a function of $F$, $q_0$, $|q|$.  Each of the sub-plots
shows the density at fixed $F$ with abscissa $|q|$ and
ordinate $q_0$.  These sub plots are arranged with flux increasing
from left to right with flux $F$ increasing by a factor 1.43 at
each step.  All particles detected with significance level greater
than 4-sigma are shown.  The panel below this shows the density of
particles weighted by the individual object polarizability, and
the panel below that shows the density weighted by the effective
polarizability given  by (\ref{eq:Peff3}).
The lower plot shows the density of objects weighted
by the actual polarization (divided by the shear applied in the
simulation, or $\gamma = 0.1$ in this case).  
}
\label{fig:Pplot}
\end{figure}

\subsection{Optimal Weighting with Flux, Size and Shape}
\label{sec:optimisation}

Armed with the conditional mean polarization
$\langle q_\alpha \rangle_{F, q_0, q^2}$ we can now compute an
optimal weight (as a function of, for example, 
flux $F$, size $q_0$, and eccentricity $q^2)$ for combining the
estimates of the shear from galaxies of different types.
Let us assume that one has measured fluxes etc.~for a very large
number of galaxies --- from an entire survey say --- and
that from these data one has determined the mean number density
of galaxies $n(F, q_0, q^2)$ and also the various
conditional averages and gradients
that appear in (\ref{eq:Peff3}).  Now consider
a relatively small spatial subsample 
of these galaxies and bin these into cells in $F, q_0, q^2$
space, and for each bin compute the occupation number $N$ and
the summed polarization $\sum q_\alpha$. 

A shear estimator for a cell which
happens to have occupation number $N$ is
\begineq
\hat \gamma_\beta = \Peff^{-1}_{\alpha \beta} \sum q_\alpha / N.
\endeq
To simplify matters, let us neglect for now any anisotropy of the
point spread function, in which case we can write 
$\Peff_{\alpha \beta} = \Peff \delta_{\alpha \beta}$ where $\Peff \equiv \Peff_{\eta\eta} / 2$
and we then have
\begineq
\hat \gamma_\beta = {1\over N \Peff} \sum q_\beta .
\endeq
The expectation value for the variance in $\hat \gamma$ (for a cell which
happens to contain $N$ galaxies) is
\begineq
\langle \hat \gamma^2 \rangle =
\langle \sum q^2 \rangle / (N^2 \Peff^2) = q^2 / N\Peff^2
\endeq
where we have used $\langle (\sum q_\alpha) (\sum q_\alpha) \rangle = 
\langle \sum q^2 \rangle$ since, in the limit of weak shear, the
galaxy polarizations are uncorrelated.
As different cells give shear estimates whose fluctuations are
mutually uncorrelated, the optimal way to combine the shear estimates from
all the cells is to average them with weight per cell 
$W_{\rm cell} \propto 1 / \langle \hat \gamma^2 \rangle 
= N\Peff^2 / q^2$ to obtain a final optimized total shear estimate
\begineq
\hat \gamma_\alpha^{\rm total} = 
{\sum\limits_{\rm cells} (N \Peff^2 / q^2) ({1\over N \Peff} \sum q_\alpha ) \over
\sum\limits_{\rm cells} N\Peff^2 / q^2}
= {\sum\limits_{\rm galaxies} \Peff q_\alpha / q^2 \over
\sum\limits_{\rm galaxies} \Peff^2 / q^2}
= {\sum\limits_{\rm galaxies} Q \hat q_\alpha \over
\sum\limits_{\rm galaxies} Q^2}
\endeq
where $Q \equiv \Peff / q$.
Thus the optimized cell weighting scheme corresponds to averaging the
shear estimates from individual galaxies
$\hat \gamma_\alpha^{\rm galaxy} = q_\alpha / \Peff$
(for galaxies of given $F, q_0, q^2$)
with weights $W_{\rm galaxy} = \Peff^2 / q^2$.  
\hide{Note that while the
estimator (....) for the shear for an individual cell appears to be ill defined
for empty cells, the final result is well defined since the empty cells
receive zero weight.}

The variance in the total shear estimator is 
\begineq
\label{eq:totalvariance}
\langle \gamma^2 \rangle = 
{\sum Q^2 \langle \hat q^2 \rangle \over (\sum Q^2)^2} =
(\sum Q^2)^{-1}.
\endeq
The quantity $\sum Q^2$ is extensive with the number of galaxies and
its value, per unit solid angle of sky, provides a useful figure
of merit for weak lensing data. From a 2.75hr I-band integrations 
of solid angle $d\Omega = 0.165$ square degrees at CFHT taken
in good seeing ($0''.60$ FWHM) we obtained $\sum Q^2 \simeq 4.7\times 10^4$
\cite{kwl+98} or
$\sum Q^2 / d \Omega \simeq 2.85 \times 10^5 / {\rm sq\; degree}$, so
with data of this quality, the statistical
uncertainty in the net shear (per component) measured over one square degree would
be around $\sigma_\gamma \simeq (2 \times 2.85 \times 10^5)^{-1/2}
= 1.32 \times 10^{-3}$.

This figure of merit allows one to tune the parameters of one's shape measurement
scheme, such as the weight function scale size, in an unbiased and objective
manner.
The weighting scheme derived above is appropriate if the shear is independent
of the measured flux etc.  This is the case for lensing by low redshift
clusters where for all relevant values of source redshift the
sources are effectively `at infinity' and the shear has saturated at its value for
an infinitely distant source $\gamma^{\infty}_\alpha$.  For high
redshift lenses the shear will vary with source redshift, and if
one has some, perhaps probabilistic, distance information
at one's disposal, then the weighting scheme should be modified.
Let us assume that the measured photometric properties $p_i$ of some
object indicate it has a probability distribution to be at
distance $z$ of $p(z|p_i)$; with high resolution spectroscopy this would be
a delta-function at the measured redshift whereas with only broadband
colors the conditional probability would be smeared out and perhaps
multimodal.  The conditional mean shear for this object, for a given
a foreground lens, is proportional to the mean inverse critical
critical surface density, and the optimal estimate for the
shear at infinity is 
\begineq
\hat \gamma_\beta^\infty = {1 \over \Sigma_{\rm crit}(\infty)}
{\sum\limits_{\rm galaxies} \langle 1 / \Sigma_{\rm crit}(z) \rangle  Q q_\alpha
\over \sum\limits_{\rm galaxies} \langle 1 / \Sigma_{\rm crit}(z) \rangle^2  Q^2 }
\endeq

\section{Correction for PSF Anisotropy}
\label{sec:anisotropy}

In the foregoing we have computed how the shape polarization
$q_\alpha$ responds to a shear. This allows us to correctly
calibrate the circularizing effect of seeing.  In general, asymmetry of the
PSF will also introduce spurious systematic
shape polarization $\langle q_\alpha \rangle$ in (\ref{eq:gammaestimator}), and it is
crucial that this be measured and corrected for.
As discussed in \S\ref{sec:introduction}, for unweighted second moments,
the effect of PSF anisotropy is rather simple since the 
(flux normalized) final second moment $q_{lm}$ is the sum of the
intrinsic second moment and that of the instrument, so the anisotropic
parts of the instrumental second moment $q_\alpha = M_{\alpha l m} q_{lm}$
can be measured from stars, and subtracted from those observed.
For weighted or isophotal moments things are more complicated, and depend on how the
anisotropy is generated. KSB considered
a simple model in which the final PSF is the convolution of some
circularly symmetric PSF with a small, but highly asymmetric
`anisotropizing kernel' $k(r)$. 
In \S\ref{subsec:convolutionmodel} we will further explore this
`convolution model', the attempts that have been made to implement
this perturbative approach, and estimate the error in this method.
In \S\ref{subsec:recircularisation} we show that the convolution
model is quite inappropriate when applied to diffraction limited seeing
and we develop a more general technique for correcting PSF anisotropy.
Finally, in \S\ref{subsec:noisebias}
 we draw attention to a type of noise related bias that affects
shear estimators, but which has hitherto been overlooked, and
we show how this can be dealt with.

The artificial polarization produced by PSF anisotropy is an effect
which is present in the absence of any gravitational shear. Thus at
leading order in $\gamma$ we can set $\gamma = 0$.  This effectively
decouples the computation and correction of the PSF anisotropy
from the shear-polarization calibration problem considered above.
It also means that in this section we can assume that the intrinsic shapes of
galaxies are statistically isotropic.

\subsection{The Convolution Model}
\label{subsec:convolutionmodel}

KSB computed the effect of a PSF anisotropy under the assumption
that this can be modeled as the convolution of a perfect circular
PSF with some compact kernel $k(r)$.
This would be a good model, for instance, for
observations primarily limited by atmospheric turbulence
with seeing disk of radius $r_g$, 
but with very small guiding or registrations errors or optical
aberrations with extent
$\delta r \ll r_g$, and the KSB analysis gives a correction to
lowest order in $\delta r$.  
In this model, the 
effect of `switching on' the anisotropy is the
transformation on the observed image, which is smooth on scale $\sim r_g$,
\begineq
\label{eq:fobstaylorexpansion}
\fobs(r) \rightarrow \fobs'(r) = \int d^2 r' \; k(r') \fobs(r - r')
= \fobs(r) - k_i \d_i \fobs(r) + {1 \over 2} k_{ij} \d_i \d_j \fobs(r) 
- {1 \over 6} k_{ijk} \d_i \d_j \d_k \fobs(r) + \ldots
\endeq
where we have Taylor expanded $\fobs(r - r')$ and where
$k_i \equiv \int d^2 r r_i k(r)$, $k_{ij} \equiv \int d^2 r r_i r_j k(r)$ etc.
Taking the kernel to be centered, we have $k_i = 0$, and to lowest non-vanishing order 
the effect on the polarization is
\begineq
\label{eq:ksbpolarisability1}
q_\alpha' = \int d^2 r \; w_\alpha(r) \fobs'(r) = q_\alpha +
{1\over 2} k_{ij} \int d^2 r\; \fobs(r) \d_i \d_j w_\alpha(r) 
\endeq
where $w_\alpha \equiv \Malm w(r) r_l r_m$ and we have integrated by parts. 
The induced stellar polarization is linear in $k_{ij}$ and therefore
scales as the square of the extent of the convolving kernel since
$k_{ij} \sim \delta r^2$. Performing the decomposition $k_{ij} = k_A M_{Aij}$
we find that an individual galaxy's polarization $q_\alpha$ will depend on both
the trace $k_0$ and the trace-free parts $k_\alpha$ of $k_{ij}$. The
{\sl average} induced polarization, however, only depends on $k_\alpha$
and we have $\langle q_\alpha \rangle = k_\beta P^{\rm sm}_{\alpha \beta} $ with
\begineq
\label{eq:ksbpolarisability2}
 P^{\rm sm}_{\alpha \beta} = \half M_{\alpha l m} M_{\beta ij} 
\int d^2 r\; \langle \fobs \rangle \d_i \d_j (w(r) r_l r_m).
\endeq
To lowest order in the PSF anisotropy we can use the KSB polarizability (\ref{eq:ksbpolarisability2})
to infer $k_\alpha$ from the shapes of stars, and then correct the weighted second moments 
or the galaxies. At the same level of precision
one can convolve one's image with a small kernel designed to nullify
the anisotropy.  \citeN{ft97} have presented a $3 \times 3$ smoothing kernel
which does this.  In the typical situation, the shapes are measured from an average of
numerous images taken with a pattern of shifts, and which are co-registered to fractional
pixel precision, and a simpler but equally effective approach is then to average over pairs of
images which have been deliberately displaced from the true solution
by the small displacement vector 
$\pm \delta r = \sqrt{|k_A|}\{\cos(\theta), \sin(\theta)\}$ with $\theta = \tan^{-1}(k_2/k_1)/2 + \pi / 2$.

What is the error in this linearized approximation? To answer this we must
consider higher terms in the expansion (\ref{eq:fobstaylorexpansion}).
The next order correction to $q_\alpha$ involves $k_{ijk}$ which,
unlike the centroid $k_i$ cannot be set to zero, so for a given
object the fractional error in the KSB correction is on the order of
$\delta r / r_0 \sim \sqrt{e_\alpha}$. Galaxies, however,
are randomly oriented on the sky, 
so the average change in $q_\alpha$ for a galaxy of
some arbitrary morphology but averaged over all position angles 
at this order in $\delta r / r_0$ is
\begineq
\langle \delta q_\alpha \rangle \sim 
k_{ijk} \int d^2 r\; \langle \fobs(r) \rangle \d_i \d_j \d_k w_\alpha(r).
\endeq
This vanishes since $\langle \fobs(r) \rangle$ is an even
function while $\d_i \d_j \d_k w_\alpha(r)$ is odd, provided we take $w(r)$ to be
circularly symmetric at least, which we will assume is the case.
The effective net fractional error in the
KSB approximation is therefore on the order of $(\delta r / r_0)^2$, or
typically on the order of the induced stellar ellipticity.

\subsection{General PSF Anisotropy Correction}
\label{subsec:recircularisation}

\begin{figure}[htbp!]
\centering\epsfig{file=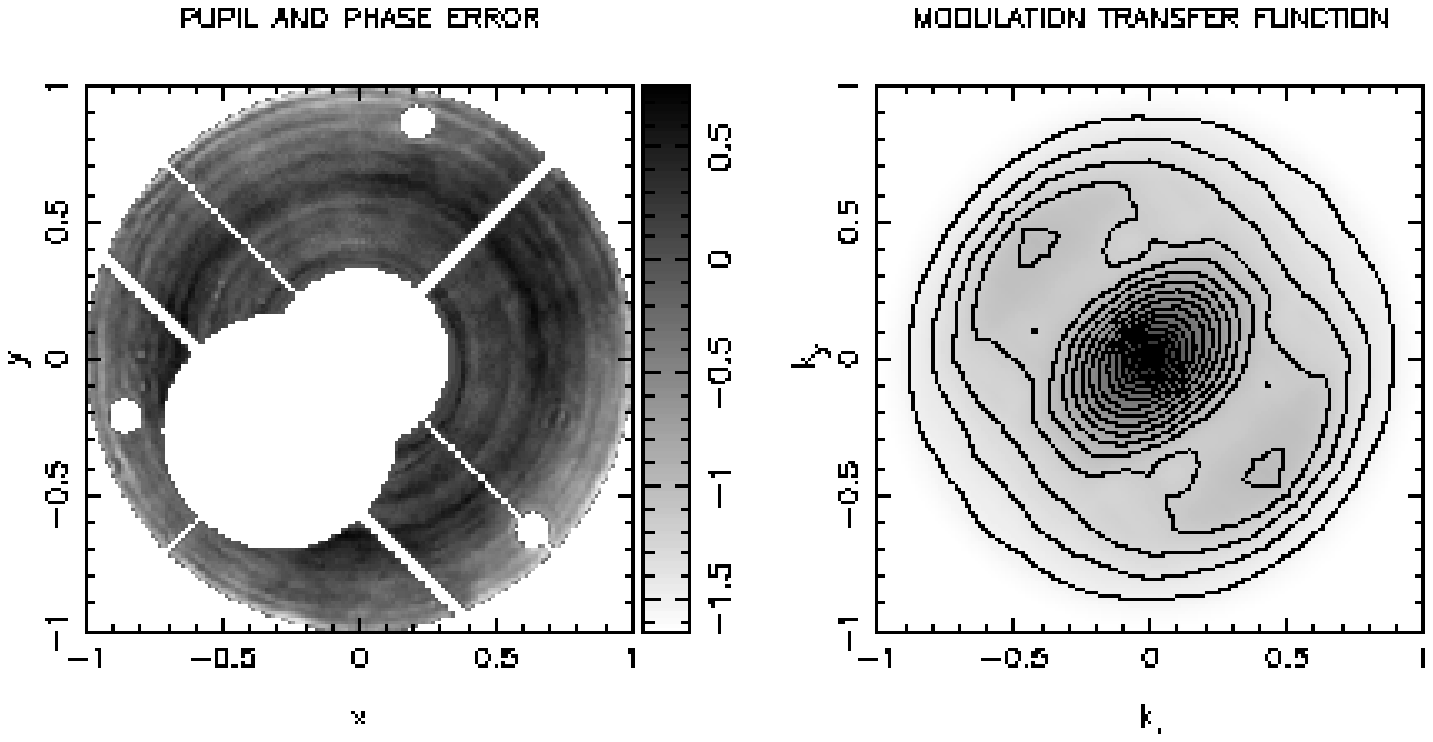,width={1.0 \linewidth},angle=0}
\figcap{The panel on left shows the WFPC2 pupil function and
phase error from Krist's Tiny Tim model for an off-axis point on the
focal plane.
The phase is given in radians for a wavelength of 800 nm.
The panel on the right shows the modulation transfer function
(the real part of the OTF).
}
\label{fig:wfpc2pupilandphase}
\end{figure}

\begin{figure}[htbp!]
\centering\epsfig{file=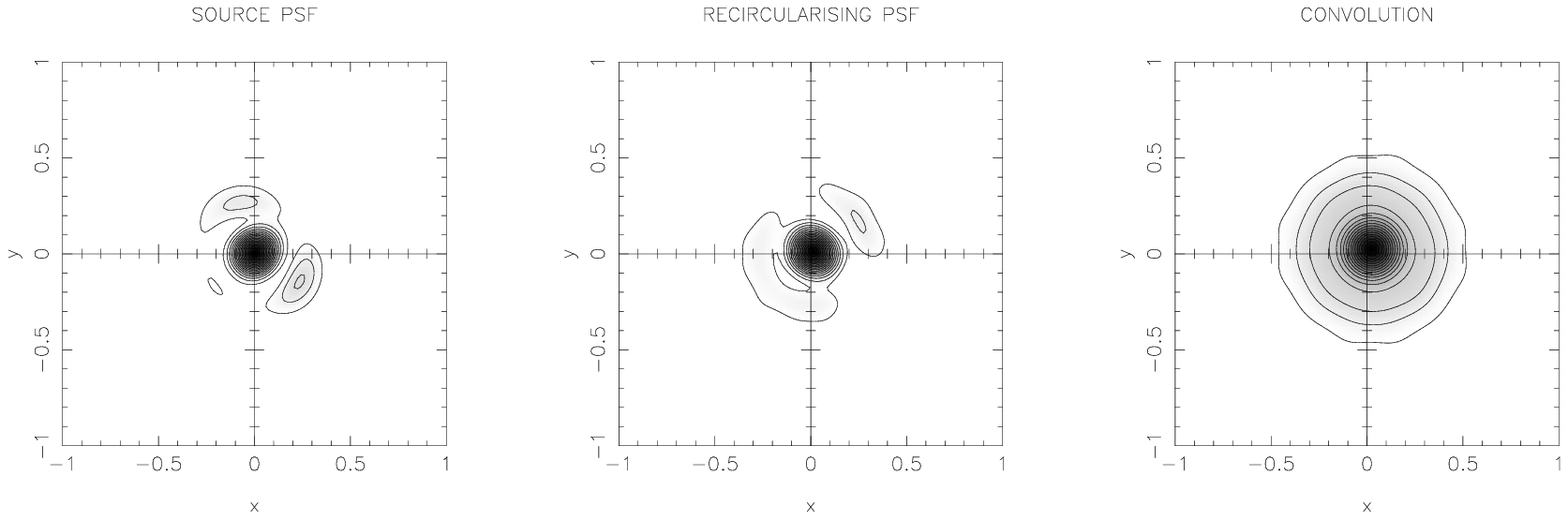,width={1.0 \linewidth},angle=0}
\figcap{HST WFPC2 PSF from Tiny Tim and re-circularizing filter
computed as described in the text.}
\label{fig:gcircplot}
\end{figure}

In principle, one can develop a higher order correction 
scheme within the context of this model, but this 
does not seem to be particularly promising; it is not clear that
the result will be sufficiently robust and accurate for even for ground based
observations --- in very good seeing conditions the PSF anisotropy
from aberrations
becomes large, and the perturbative approach will break down.
Moreover, for observations with telescopes in space, the model of the
PSF as a convolution of a perfect circular PSF with a
kernel, small or otherwise, is wholly unfounded.
The instantaneous OTF is
\begineq
\label{eq:diffotf}
\tilde g(k) = \int d^2r \; A(r) A(r + k D \lambda / 2 \pi)
\exp(i[\varphi(r) - \varphi(r + k D \lambda / 2 \pi)])
\endeq 
where $A$ is the real transmission function of the telescope input pupil,
$D$ is the focal length, and
$\varphi(r)$ is the phase error due to mirror aberrations and the atmosphere.
It is often said that the general OTF factorizes into a set of terms describing
the atmosphere; the telescope aperture; and aberrations, and this would
seem to justify the convolution model discussed above.  For atmospheric 
turbulence, and for aberrations arising from random
small scale mirror roughness this is correct.  This is because the
average of the complex exponential term in (\ref{eq:diffotf}) depends only on
$\delta r = k D \lambda / 2 \pi$ and is independent of $r$, and
(\ref{eq:diffotf}) then factorizes into two independent terms, and
the PSF is then the convolution of  two completely independent and
non-negative functions, but this is not the case in general.
To be sure, one can write the combined aperture and
aberration OTF as a product of some `perfect' OTF with some
other function (the true OTF divided by the perfect one) but quite unlike the
case for random turbulence and mirror roughness, this function
is neither independent of the shape of the pupil, nor is it
positive; if you compute this function for a telescope subject to a 
low order aberration from figure error, for instance, you will find that the this
function is strongly oscillating and is just as extended as the true PSF.
Consider the situation 
in WFPC2 observations, for example, where the 
PSF anisotropy has important contributions from both 
from asymmetry of the pupil $A(r)$ and from phase errors,
though with the former tending to dominate (though not enormously so) for long wavelengths
and far off axis with the WFPC2. 
The aperture function for the lower left corner of chip 2 computed from
the Tiny-Tim model
\cite{krist95} is shown in  figure \ref{fig:wfpc2pupilandphase}.
The off-axis pupil function is approximately that on-axis minus
a disk of some radius $r_1$
at some distance $\Delta r$ off-axis.  If we let the radius of the
primary be $r_0$ and define a disk function
$D_{r_0}(r) = \Theta(r / r_0 - 1)$ then on computing
the electric field amplitude $a(x)$ and
squaring we find that
the on-axis PSF is of course given by the Airy disk: $g \simeq \tilde D_{r_0}^2$
while the off axis PSF $g'$ is given by
$g'(x) \simeq g(x) - \tilde D_{r_1} \tilde D_{r_0} \cos(2 \pi x \Delta r / D \lambda)$.
For $r_1 \ll r_0$, $\tilde D_{r_1}$ is relatively slowly varying and close to unity,
in which case
the extra off-axis obscuration introduces a perturbation which is 
proportional to the
product of $D_0$ and a planar wave, or essentially an asymmetric modulation of the
side lobes of the on-axis PSF (see LH panel of figure \ref{fig:gcircplot}).   

In general then, and especially for diffraction limited observations,
the perturbative convolution model cannot be trusted.
Luckily, at least in the context of the shear estimators discussed
here, a simple solution clearly presents itself:
Since to compute the polarizability requires that
one generate a rather detailed model of the PSF, and that
we probably want to apply some kind of
re-convolution $\fobs \rightarrow f_s = \gdag \otimes \fobs$
to obtain a well behaved shear response, we might
as well use the opportunity to choose $\gdag(r)$ in order to
re-circularize the PSF exactly.  
For example, from the observed PSF, one can compute the
OTF $\tilde g(k)$ and then form the circularly symmetric function $\tilde g_{\rm min}(k)$ 
being the greatest real function which lies everywhere below $|\tilde g(k)|$.
The function $\gdag(r)$ with
transform
\begineq
\tilde \gdag(k) =  \tilde g_{\rm min}(k)^2 / \tilde g(k)
\endeq
both guarantees a well defined shear polarizability and gives
a perfectly circular resulting total PSF.
This is illustrated in figure \ref{fig:gcircplot}.

In some situations the PSF may have near 180 degree symmetry, in
which case a good approximation
is to re-convolve with a PSF which has been rotated through 90 degrees.
Since the PSF is real, the OTF must satisfy the symmetry
$\tilde g(k) = \tilde g^*(-k)$. 
In the absence of aberrations, a diffraction
limited telescope has OTF $\tilde g = A \oplus A$
which is real and so has exact symmetry under rotation by $\pi$,
a property which is shared the PSF, so $R_\pi g = g$.
If one convolves with $\gdag = R_{\pi/2} g$ then the
resulting total PSF $\gdag \otimes g$ is symmetric under 90 degree
rotations (proof: $R_{\pi/2} \tilde g_{\rm tot} = 
R_{\pi/2} (R_{\pi/2} \tilde g \tilde g) = R_\pi \tilde g
R_{\pi/2} \tilde g = \tilde g R_{\pi/2} \tilde g = \tilde g_{\rm tot}$).
For a galaxy of given form, the average PSF induced polarization is
\begineq
\langle q_\alpha \rangle =
\int d^2 r \; \langle g_{\rm tot} \otimes f \rangle w_\alpha =
[\langle g_{\rm tot} \otimes f\rangle \otimes w_\alpha]_{r = 0}
= \int {d^2 k \over (2 \pi)^2} \langle \tilde f \rangle 
\tilde g_{\rm tot} \tilde w_\alpha
\endeq 
where $w_\alpha = M_{\alpha l m} w(r) r_l r_m$.
But $ \langle \tilde f \rangle $ is circularly symmetric and
$\tilde w_\alpha$ is anti-symmetric under 90 degree
rotations, so  $\langle q_\alpha \rangle = 0$.
Other situations where this would work would be ground
based observations but with large amplitude telescope oscillations,
and in fast guiding.
In general, however, aberrations from figure errors will introduce both real
and imaginary contributions to the OTF and the latter will
destroy the exact 180 degree symmetry of the PSF.  This is
certainly the case for the WFPC2 PSF at optical wavelengths.

In the linearized approximate schemes described earlier, the PSF anisotropy
is entirely characterized by the two coefficients $k_\alpha$. In general,
these will vary substantially with position on the image, but
this can be treated by modeling $k_\alpha$ as some smoothly
varying function of image position $R$ such as a low order polynomial.
In the
scheme proposed here we need to be able to generate not just these
two coefficients, but the full two dimensional
PSF $g(r, R)$ (being the
intensity at distance $r$ from the centroid of a star at position $R$).
A simple practical approach is to solve for a model
$g(r,R) = \sum_I g_I(r) f_I(R)$, with $f_I(R)$ being some
set of polynomial or other basis functions.  
A least squares fit for the image valued mode coefficients is
$g_I(r)  = m_{IJ}^{-1} G_J(r)$ with
$m_{IJ} = \sum_{\rm stars} f_I(R) f_J(R)$; $G_I(r) = \sum_{\rm stars} f_I(R) g_\obs(r)$.
Armed with the image valued coefficients one can then, for example,
compute the convolution $f_s = g \otimes \fobs$ as
$f_s(r) = \sum_I f_I(r) (g_I \otimes \fobs)$, that is we convolve 
the source image with each of 
components $g_I(r)$ and then combine with spatially varying coefficients
$f_I(R)$.  
\hide{This is only valid if the scale length for changes of the PSF greatly
exceeds its own width, but that is usually an excellent approximation.}
Non-linear functions of the PSF such as the 
re-circularizing kernel
$\gdag$ are easily generated by making realizations of the PSF models
on say a coarse grid of points covering the source image, and from each of these computing the
desired function, and then fitting the results to a low order polynomial
just as for $g(r,R)$.

The approach described here lends itself very nicely to observations
with large mosaic CCD cameras, in which misalignment of the chip surfaces
and the focal plane coupled with telescope aberrations can give rise to
a PSF which varies smoothly across each chip, but which changes discontinuously
across the chip boundaries.  This results in a very complicated PSF pattern
on the final image obtained by averaging over many dithered images. It
is a good deal simpler to re-circularize each of the contributing
images.  We still need to generate a spatially varying model of the
final $g(r)$, $\gdag(r)$ in order to compute the polarizability, but 
if we fail to model these exactly, it will only introduce a relatively
minor error in the shear-polarization calibration.

\subsection{Noise Bias}
\label{subsec:noisebias}

In the absence of noise, the recircularization procedure exactly annuls any PSF
anisotropy; the {\sl signal\/} content of the image is exactly statistically isotropic.
The {\sl noise\/} component in the images (which in the original images is incoherent
Poisson noise) will however be correlated with an anisotropic two-point function -- the
peaks and troughs of the noise will appear as ellipses with correlated position
angles. For a given object, the noise is equally likely to produce a positive
or negative fluctuation in the polarization $q_\alpha$, and the effects
cancel out on average. Due to the nature of faint galaxy counts, however, objects of a 
given observed flux are more likely to be intrinsically fainter objects which have been
scattered upward in flux than intrinsically brighter objects which have been
scattered down, and there will therefore be
a tendency for faint objects to be aligned like the 
re-circularizing PSF; i.e.~oriented opposite to the PSF. We will refer to
this as `noise-bias'.
It should noted that an analogous effect is present in the old
KSB anisotropy correction scheme. In that case the noise is isotropic, but
objects are detected as peaks of a smoothed image and there is then
a tendency for the galaxies close to the threshold for detection to be
aligned like the PSF.  In the context of perturbative schemes as described
in \S\ref{subsec:convolutionmodel} there are other unevaluated 
errors in the correction which are typically of similar order
to the noise bias effect, which goes some way to explain why it has
been ignored in the past. In the improved 
PSF anisotropy correction method proposed here
it is the leading source of error and it behooves us to analyze
and correct for it.

Noise will affect the photometric parameters like $q_\alpha$ in
two ways; in addition to a straightforward additive noise term, noise
will also affect the object detection, and will shift the centroid
about which we measure the centered second moments, for instance.
In fact, the de-centering effect is quite weak. To see why,
consider the following simple (though actually quite practical
and realistic) model for the detection and measurement
process in which we take the
re-circularized image $f_s$ and define objects as peaks of the
field $F(r) = w \oplus f_s$ where $w$ is a some smooth, circularly symmetric
weight function (we typically use a compensated `Mexican hat' filter),
and the object moments $q_A$ are the value of the field
$q_A(r) = w_A \oplus f_s$ evaluated at the peak location.
Let us assume that there exists some object which, in the absence of noise,
would lie at the origin, so for this object we would have
\begineq
\begin{matrix}{
F = F(0) = \int d^2r \; w(r) f_s(r) \cr
0 = d_i(0) = \int d^2r \; (\d_i w) f_s(r) \cr
q_A = q_A(0) = \int d^2r \; w_A(r) f_s(r)
}\end{matrix}
\endeq
where the condition that the object be a peak of $F$
is expressed as the vanishing of the dipole-like quantity
$d_i(r) = \d_i F$.  
If we now add noise, and let $f_s \rightarrow f_s + f_n$,
where $f_n(r)$ is
the noise component of the re-circularized image,
all of the photometric fields
$F(r)$, $d_i(r)$, $q_A(r)$ will change to
$F' = F + \delta F = F + w \oplus f_n$ etc.  The dipole
$d_i$ will no longer vanish at the origin, but will have some
value $d_i(0) = \int d^2r\; (\d_i w) f_n$, and the
peak of $F(r)$ will have moved to some position $r_{\rm pk} \ne 0$.
In the vicinity of the peak we have
$d_i(r) \simeq 0 + (r - r_{\rm pk})_j \times 
\d_j d_i(r)$ and hence, to first order in the
noise amplitude, the peak location is
\begineq
r_{i,\rm pk} = - (\d_j d_i)^{-1} \delta d_j
= - \left[\int d^2 r\; \d_i \d_j w f_s \right]^{-1} 
\int d^2r\; \d_j w f_n
\endeq
The shift of the peak has no effect on $F$ at first order (because
$F$ is stationary) but is does affect the central moments $q_A$ and we have
\begineq
\label{eq:deltaFqnoise}
\begin{matrix}{
\delta F = \int d^2r \; w(r) f_n(r) \cr
\delta q_A = \int d^2 r 
\left[w_A(r) -  v_{Aj} \d_j w\right] f_n
}\end{matrix}
\endeq
where
\begineq
v_{Ai} = \left[ \int d^2r \; (\d_i \d_j w) f_s \right]^{-1}
\int d^2 r \; (\d_i w_A) f_s
\endeq
Thus, in general, the first order change in the second moments $q_A$
depends not just on the noise and the kernel $w_A$, but on the
form of the underlying noise-free image $f_s$ through the term involving $v_{Aj}$.  
There is good reason to think
that this form dependence is weak for real objects; the function
$\d_i w_A$ is an odd function, so the extra term vanishes if the galaxy
is symmetric under 180 degree rotations. Also, it is hard to see
why the presence of this term would cause a systematic polarization; the
change in the polarization $\delta q_A$ associated with the centroid
shift is proportional to the vector $v_{Aj}$ which, being a function of the
re-circularised image $f_s$ is equally likely to be positive or negative.
For each galaxy that, for a given realisation of noise, suffers a
certain $\delta q_A$ there is a 180 degree rotated clone which has
precisely the same weighted flux, polarization etc.~(these being
even functions) but which suffers a  $\delta q_A$ of opposite sign.
The kind of effect envisioned in the leading paragraph of this sub-section
arises if there is a correlation between fluctuations in the polarisation
and the flux $F$, coupled with a gradient of the density of objects
$n(F)$.  The first term in $\delta q_A$ in (\ref{eq:deltaFqnoise})
does, as we shall see, indeed correlate with 
$\delta F$, but, when we average over orientation of the underlying
noise-free galaxies, the second does not. Unfortunately, we
have not been able to rigorously demonstrate that the
centroid shift term vanishes. Nonetheless,
in what follows
we will assume that
this term is negligible.  This should certainly be adequate in order to
estimate the magnitude of the effect, and is probably sufficiently accurate to
give a useful correction, though the results should strictly be
regarded as only approximate.

Let us now assume one has measured a set of $n$ photometric parameters
$p_i$ for a galaxy which, like $F$, $q_0$, $q_\alpha$, are linear 
functions of the
brightness $\fobs$ and of the form $p_i = \int d^2r \; K_i(r) \fobs(r)$.
The effect of noise in the images will be to introduce
a perturbation in these parameters $\delta p_i$.  These perturbations have
zero mean, $\langle \delta p_i \rangle = 0$, and
since there are typically
a very large number of photons (noise plus signal) contributing to the
parameters $p_i$, the central limit theorem dictates that the $\delta p_i$
will have a multivariate Gaussian distribution:
\begineq
\sigma(\delta p) d^n\delta p = ((2 \pi)^n |C|)^{-1/2} 
\exp(- \delta p_i C^{-1}_{ij}  \delta p_j / 2)
d^n \delta p
\endeq
where the covariance matrix is $C_{ij} = \langle \delta p_i \delta p_j \rangle$
and is given by
\begineq
\label{eq:covariance}
C_{ij} = \int d^2 r \; K_i (K_j \otimes \xi_n)
\endeq
where $\xi_n(r) = \langle f_n(r') f_n(r' + r) \rangle$ is the 2-point
function of the noise $f_n(r)$,
and which may be computed as described in appendix \ref{sec:PSFtheory}.

For objects which are detected at a fairly high level of significance $\nu \gg 1$,
the noise will cause a small modification of the underlying distribution
function: $n(p) \rightarrow n'(p)$, 
which we can compute in a perturbative manner:
\begineq
n'(p) = \int d^n \delta p \; n(p - \delta p) \sigma(\delta p)
= n(p) - {\d n \over \d p_i} \langle \delta p_i \rangle
+ \half {\d^2 n \over \d p_i \d p_j} \langle \delta p_i \delta p_j \rangle
+ \ldots
\endeq
Since $\sigma(p)$ is an even function, all the odd terms in this expansion
vanish, and we have $n'(p) = n(p) + \half C_{ij} \d^2 n / \d p_i \d p_j$
at leading order.

Let us specialize now to the $n=4$ dimensional case: $p_i = F, q_0, q_\alpha$
and compute the mean polarization for galaxies of given $F$, $q_0$
with some weight function $W(q = |q_\alpha|)$: 
\begineq
\langle q_\alpha \rangle_{F, q_0} = 
{\int d^2q \; W(q) q_\alpha n'(F, q_0, q_\alpha) 
\over \int d^2q \; W(q) n'(F, q_0, q_\alpha)} \simeq
{\int d^2q \; W(q) q_\alpha \delta n(F, q_0, q_\alpha) 
\over \int d^2q \; W(q) n(F, q_0, q_\alpha)}
\endeq
where $\delta n(F, q_0, q_\alpha) = \half C_{ij} \d^2 n / \d p_i \d p_j$.  Since $q_\alpha$ is an
odd function and $W(q)$, $n(F, q_0, q_\alpha)$ are even,
the only terms which contribute to the integral in the numerator are
those involving a single derivative with respect to polarization: 
$\d^2 n /\d F \d q_\beta$ and $\d^2 n /\d q_0 \d q_\beta$ and, 
on integrating by parts, we have
\begineq
\langle q_\alpha \rangle = 
{- \half \int d^2q \; 
\left(\langle \delta F \delta q_\beta \rangle {\d n \over \d F} +
\langle \delta q_0 \delta q_\beta \rangle {\d n \over \d q_0}\right)
{\d W(q) q_\alpha \over \d q_\beta} 
\over \int d^2q \; W(q) n(F, q_0, q_\alpha)}
\endeq
Using 
$\d(W(q) q_\alpha) / \d q_\beta = \delta_{\alpha \beta} W(q) 
+ q_\alpha \hat q_\beta dW / dq$, integrating over angle
and converting from $n(F, q_0, q_\alpha)$ to
$n'(F, q_0, q) = 2 \pi q n(F, q_0, q_\alpha)$ and dropping the prime we have
\begineq
\langle q_\alpha \rangle = 
{-\half \int dq \;
\left(\langle \delta F \delta q_\alpha \rangle {\d n \over \d F} +
\langle \delta q_0 \delta q_\alpha \rangle {\d n \over \d q_0}\right)
(W(q) + \half q W'(q))  
\over \int dq \; n W(q)}
\endeq
and letting $W(q)$ become a delta-function to isolate a single value of $q$ we have
\begineq
\label{eq:noisebias}
n \langle q_\alpha \rangle = - \half 
\left(\langle \delta F \delta q_\alpha \rangle {\d n \over \d F} +
\langle \delta q_0 \delta q_\alpha \rangle {\d n \over \d q_0}\right)
\endeq
Thus, as anticipated, there is a net induced polarization if
there is a non-zero correlation between the polarization fluctuation
$\delta q_\alpha$ and $F$ and/or $q_0$, and also significant gradients
of the distribution function with respect to $F$ or $q_0$.

The effect 
on the re-scaled polarization of the first term in (\ref{eq:noisebias}) is
$\langle q'_\alpha \rangle =
\langle q_\alpha \rangle / F = -(\half \langle \delta F \delta q_\alpha \rangle
/ F^2)
\d \ln n / \d /ln F$ and
is second
order in the inverse significance: 
$\langle q'_\alpha \rangle \propto \nu^{-2}$, where $\nu^2 = F^2 / \langle \delta F^2 \rangle$ and
rapidly becomes small for well observed objects.
Thus it should be possible to set a sensible limit on the significance
of object detection, and, if necessary,  use
(\ref{eq:noisebias}) together with (\ref{eq:covariance}) for
$\langle \delta F \delta q_\alpha \rangle$ etc.~to correct for this.
The error in
this linearized approximation is of fourth order in the inverse
significance.
As a simple illustrative example consider a Gaussian galaxy
of scale $r_G$, a Gaussian
window function $w$ with scale $r_w$, and a Gaussian ellipsoid
PSF $g = \exp(-(x^2 / r_a^2 + y^2 / r_b^2)/2)$ with
$r_a = r_g(1 + \epsilon / 2)$, $r_b = r_g (1 - \epsilon / 2)$.
A suitable recircularising kernel is then 
$\gdag = \exp(-(x^2 / r_b^2 + y^2 / r_a^2)/2)$ and
the normalised two point function of the noise is then
\begineq
\xi_n(r) = {1 \over \pi r_a r_b} e^{-(x^2 / 2 r_b^2 + y^2 / 2 r_a^2)/2} 
\endeq 
The expectation averages are, to first order in $\epsilon$,
$\langle \delta F^2 \rangle = 2 (r_w^2 + r_g^2)$;
$\langle \delta q_1 \delta F \rangle = r_w^4 (r_a^2 - r_b^2) / (r_w^2 + r_g^2)
= 2 \epsilon r_w^4 r_g^2 / (r_w^2 + r_g^2)$, and so
taking only the first term in (\ref{eq:noisebias})
for simplicity
\begineq
\langle \delta q_1' \rangle  
\equiv {\langle \delta q_1 \rangle \over F} 
= - \half
{\d \ln n \over \d \ln F} {1 \over \nu^2} 
{\langle \delta q_1 \delta F \rangle \over \langle \delta F^2 \rangle}
\simeq \half {\epsilon \over \nu^2} r_w^4 r_g^2 / (r_w^2 + r_g^2)
\endeq
using $\d \ln n / \d \ln F \simeq 1$ as observed.
A shear of strength $\gamma$ applied to the Gaussian galaxy
produces a polarisation
\begineq
\langle \delta q_1' \rangle = 4 \gamma r_w^4 r_G^2 / (r_G^2 + r_w^2 + r_g^2)
\endeq
therefore a PSF anisotropy of strength $\epsilon$ is equivalent 
to an effective shear
\begineq
\gamma = {\epsilon \over 8 \nu^2} {r_g^2 \over r_G^2}
{(r_w^2 + r_g^2 + r_G^2)^2 \over (r_w^2 + r_g^2)^2}
\endeq
or, for $r_w = r_G = \sqrt{2} r_g$ say, $\gamma = 25 \epsilon / 144 \nu^2$, which for
$\nu = 6$ and PSF asymmetry $\epsilon = 0.3$, which is a reasonable value
for off-axis points on the CFHT in good seeing,
gives a shear of around $1.4\%$.
This is a sizable effect, and should therefore be corrected for.

\section{Discussion}
\label{sec:discussion}

We have considered the problem of how to estimate weak gravitational
shear from observations which have been degraded by atmospheric and/or
instrumental effects.  Previous analyses of this problem have made
simplifying assumptions which render the results inaccurate.
A major result of the paper is the finite resolution
shear operator (\ref{eq:fobsoperator}) which gives the response of
an observed image to a gravitational shear applied before smearing with
the PSF.  This result can be used to properly calibrate the effect of
any shear estimator, and is valid for arbitrary PSF, be it turbulence 
or diffraction limited.  
We then focused on the application to
weighted moment shear estimators.  We have computed the response of
individual objects to a shear in \S\ref{subsec:individualresponse},
and the response of the population of background galaxies 
with given photometric properties in \S\ref{subsec:populationresponse}, and
from this we have devised an optimal weighting scheme
\S\ref{sec:optimisation}.
In the last section we have considered the correction for
PSF anisotropy.  While there are still some approximations in the
present analysis, we feel that they place the techniques of
shear measurement on a much firmer footing than before.

\section{Acknowledgements}
I wish to acknowledge helpful discussions with
Malcolm Northcott, Francois and Claude Roddier, John Tonry, Ger Luppino,
Pat Henry, Ken Chambers, Jeff Kuhn and Christ Ftaclas.


\appendix

\section{Properties of Optical Point Spread Functions}
\label{sec:PSFtheory}

Here we shall briefly review and derive some properties of 
telescope point spread functions
which are used above. For more detailed background the reader should
consult (\citeNP{roddier81}; \citeNP{beckers93}) and references therein.
We will highlight
the wavelength or color dependence of the various 
sources of PSF anisotropy, which may
be crucially important for weak lensing searches for large-scale structure
and galaxy-galaxy lensing. 

According to elementary diffraction theory \cite{bw64} 
the complex electromagnetic field
amplitude $a(x)$ due to a distant source at position $x_{\rm phys}$ on the focal plane
(we will suppress polarization subscripts
for clarity) is given as an integral over the input pupil
\begineq
\label{eq:fresnelintegral}
a(x_{\rm phys}) = \int d^2r \; A(r) C(r) e^{2 \pi i x_{\rm phys} r / L \lambda}
\endeq
where $A(r)$ is the `pupil function' describing the aperture transmission,
$C(r)$ is the complex electric field amplitude of the incoming
wave, 
$\lambda$ is the wavelength of the radiation and $L$ is the
focal length.  
The field amplitude $C(r)$ will incorporate any random amplitude and
phase variations of the
incoming wavefronts due to atmospheric turbulence,
whereas constant wavefront distortions due to  aberrations in the optical
elements of the telescope are incorporated as a complex factor
in $A$. Thus $a(x)$ is the Fourier
transform of $AC$, evaluated at wave-number $k = 2 \pi i x / D \lambda$. The
PSF $g$ is the square of the field amplitude and
in rescaled
coordinates $x = 2 \pi x_{\rm phys} / L \lambda$, is
\begineq
\label{eq:PSF1}
g(x) = |a(x)|^2 = \int {d^2z \over (2 \pi)^2} e^{-ix\cdot z} \tilde g(z)
\endeq
where the OTF is
\begineq
\label{eq:OTF1}
\tilde g(z) = \int d^2r C(r) C^*(r+z) A(r) A^*(r + z).
\endeq
For very short ground-based observations 
the atmospheric rippling is frozen and the PSF consists of speckles. 
For long exposures we are taking the time average 
of the OTF and can replace $C(r) C^*(r+z)$ by its
time average $\langle C(r) C^*(r+z) \rangle = \xi_C(z)$. This
is independent of $r$, so the OTF factorizes into two independent
functions
\begineq
\label{eq:OTF2}
\tilde g(z) = \xi_C(z)
\int d^2 r A(r) A^*(r + z)
\endeq
and the same is true for random small scale amplitude or phase
fluctuations introduced by e.g.~random fine scale mirror roughness.

\subsection{Atmospheric Turbulence}

Ground based observations on large telescopes are usually limited by
atmospheric seeing arising from inhomogeneous random
turbulence, and it is a good
approximation to ignore the finite size of the entrance aperture and set 
the factor involving $A$ in (\ref{eq:OTF2}) to unity.
In the `near field' immediately behind the turbulent layer,
the effect on the incoming wave
is a pure phase shift
$C(r) = e^{i\varphi(r)}$ 
where $\varphi = 2 \pi i d(r) / \lambda$ and $d(r)$ is the
vertical displacement of the wavefront due to turbulence.
The displacement $d$ is nearly independent of wavelength, so
$\varphi(r) \propto 1/ \lambda$.
At greater depths this phase shift evolves into a combination of
amplitude and phase variations, but the 2-point
function $\langle C(r) C^*(r+z) \rangle$ remains invariant 
(\citeNP{fried66}; \citeNP{roddier81})and
the `natural seeing' OTF is 
\begineq
\tilde g(z) = \langle e^{i(\varphi(r) - \varphi(r + z))} \rangle.
\endeq
For steady turbulence and long integrations the central limit
theorem guarantees that the phase error $\psi =
\varphi(r) - \varphi(r + z)$ will have a Gaussian probability distribution
$p(\psi) = (2 \pi \langle \psi^2 \rangle)^{-1/2} \exp(- \psi^2 / 2 \langle \psi^2 \rangle)$
and so the time average of the complex exponential is
\begin{equation}
\label{eq:expavg}
\langle e^{i \psi} \rangle = \int d\psi \; p(\psi) e^{i \psi} =
\exp(-\langle \psi^2 \rangle / 2) = \exp(-S_\varphi(r) / 2)
\end{equation}
where the `phase structure function' is
\begineq
S_\varphi(\Delta r) \equiv \langle (\varphi_1 - \varphi_2)^2 \rangle.
\endeq
There are strong theoretical \cite{tatarski61}
and empirical reasons to believe that
on scales much less than some `outer scale' the turbulence will
have the Kolmogorov $n = -11/3$ spectrum, for which $S_d(r)  \propto r^{5/3}$.  The
structure function for the phase is conventionally written as
$S_\phi(r) = 6.88 (r/r_0)^{5/3}$ where
$r_0$ is the `Fried length' \cite{fried66} being on the order of tens of cm for
typical observing conditions (an $r_0$ of 20cm gives a FWHM = $0''.5$ at
$\lambda = 550$ nm).  The rms phase difference rises with separation
as $r^{5/6}$ in the `inertial range' delimited at the upper end by the 
outer scale, set by the width of the mixing layer, which recent
estimates (\citeNP{azb+97}; \citeNP{mtz+98}) find to be around $10-20$m
and much larger than $r_0$. The on-axis OTF computed from stars in deep CFHT imaging
agrees quite well with the theoretical expectation.
The inertial range is limited at the low end by diffusion, but at scales
much smaller than $r_0$, so little error is incurred in ignoring
this; in real telescopes mirror roughness and other effects modify the
OTF at small scales. These effects dominate the PSF at very large radii, but are
unimportant for weak lensing observations.

The optical transfer function is
$\tilde g(k) = \exp(-S_\phi(k D \lambda / 2 \pi) / 2)$ and
is real and positive. The PSF is the transform of $\exp(-6.88(z/r_0)^{5.3})$
with width which scales as $R_{\rm FWHM} \propto \lambda^{-1/5}$
which again is found to apply quite well in practice.  
This very weak dependence on
wavelength is a blessing in weak lensing since one uses stars to 
measure the PSF for the galaxies, yet the stars and the galaxies may
have different colors.
The atmospheric PSF is expected to be isotropic.
At large angles the PSF has profile $g \propto x^{-11/3}$ so the
unweighted second moment of the PSF is not well defined.
For Kolmogorov turbulence the log of the OTF is just proportional
to $k^{5/3}$ and is well defined for all $k$.

\subsection{Fast Guiding}

According to the Kolmogorov law, the rms wave-front tilt, averaged over scale
$r$ varies as $r^{-1/6}$, which suggests that even for telescopes with
reasonably large $D/r_0$ there may be useful gain in image quality
from fast guiding, and experience with HRCAM on CFHT \cite{mgr+89}
would seem to support this, though part of the dramatic improvement 
is likely due to inadequacy of the existing slow guiding system.
A technological advance which may have implications for weak lensing is the
advent of on-chip fast guiding \cite{tbs97} with OTCCD chips.  With a mosaic camera
composed of such devices it should be possible to obtain
partial image compensation over a large angular scale, 
with one or more guide stars for each `isokinetic'
patch.

The theoretical fast guiding PSF was first explored by \citeN{fried66}
who argued that the OTF should take the form of the natural seeing 
or uncompensated OTF times an `inverse Gaussian' $\exp(+\alpha k^2)$,
with scale factor $\alpha$ given in terms of $D$ and $r_0$.
This is a physically reasonable picture, since it implies that the
natural PSF is the convolution of the corrected PSF with a Gaussian
to describe the distribution of tilt,
but is only an approximate result.  Modified forms of the `Fried
approximation' have been explored by \citeN{young74} and \citeN{jenkins98}, and
the fast guiding OTF has been simulated by \citeN{christou91}.

\begin{figure}[htbp!]
\centering\epsfig{file=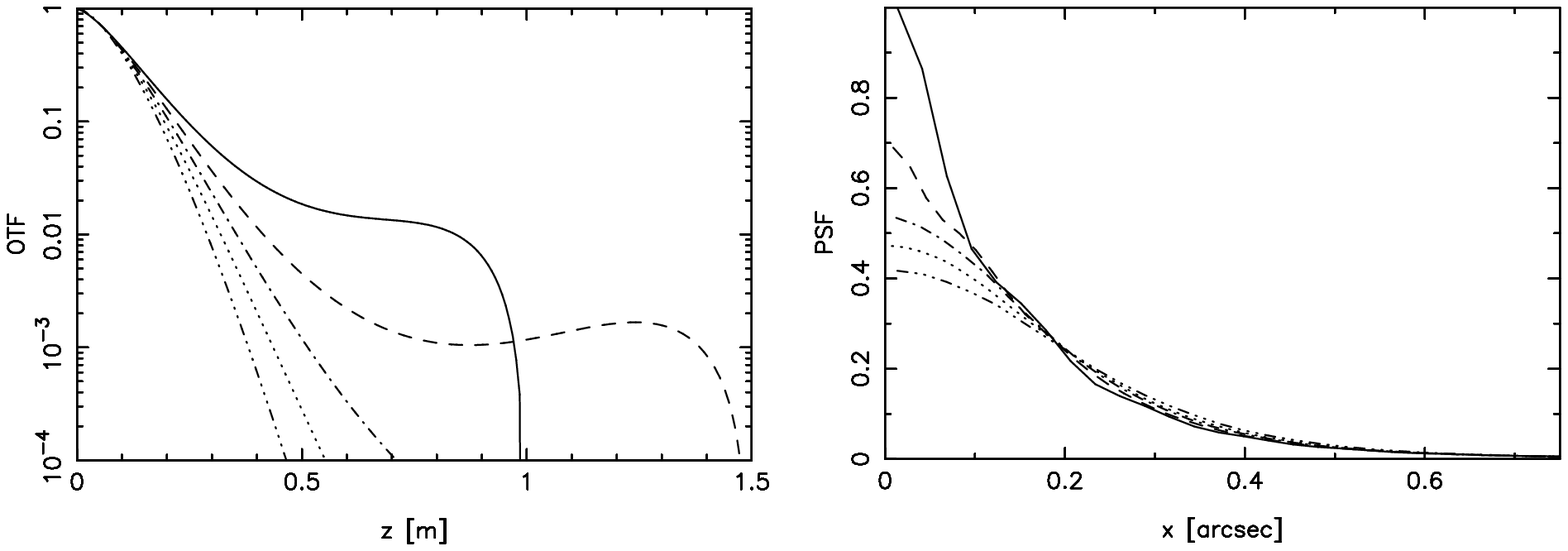,width={1.0 \linewidth},angle=0}
\figcap{Optical transfer functions and corresponding PSFs given by 
equations (\ref{eq:fastguiding1}, \ref{eq:fastguiding2})
for a range of telescope aperture diameters: 
$1.0$m solid;
$1.5$m dash;
$2.2$m dot-dash;
$3.6$m dotted;
$\infty$ dot-dot-dot-dash.
A von Karman turbulence spectrum
with outer scale of 20m and Fried length $r_0 = 0.2$m were assumed,
and the telescope was assumed to be operating at a wavelength
$\lambda = 550$nm.
}
\label{fig:fastguiding}
\end{figure}

It can be shown that in the near-field limit the exact fast-guiding OTF is given by
\begineq
\label{eq:fastguiding1}
\tilde g(z) = \int d^2 r\; A(r) A(r + z) \exp(- \langle \psi(r,z)^2 \rangle /2)
\endeq
where
\begineq
\label{eq:fastguiding2}
\langle \psi(r,z)^2 \rangle =
S(\bz) + z_i [(W_i \otimes S)_\br - (W_i \otimes S)_{\br + \bz}] 
- \half z_i z_j \int d^2r' \; W_i(\br') (W_j \otimes S)_{\br'}
\endeq
and where $W_i \equiv \d_i A^2$. Examples are shown in
figure \ref{fig:fastguiding}.  These plots show that the impact of
fast guiding on the atmospheric PSF for large telescopes will be rather
modest, at least if current, rather low, estimates of the
outer scale are correct.  Fast guiding may, however, yield dramatic improvements
for small ($\sim 1$m diameter) telescopes.
How well this would work depends largely on the altitude of the
turbulent layer.  The isokinetic patch size (over which stars
move coherently) is $\sim D / h$ so for $h = 10$km and $D=1$, say, 
this is on the order of $20''$, the motion needs to be sampled at
a rate $\gsim v / D$ reflecting the relatively high wind speed at
high altitude, 
 and it may then be difficult to
find bright enough guide stars.
There are strong indications (\citeNP{cr85}; \citeNP{tbs97}; \citeNP{mfa+91})
that centroid motions are coherent over much larger angular scales than this,
indicating that much of the image degradation arises from
low-altitude turbulence,
and this greatly improves the outlook as one can determine the
local motion by averaging a number of stars, and one can afford to sample at a lower rate.
A collection of
small telescopes equipped with wide angle OTCCD cameras could be a formidable
instrument for weak lensing or other projects requiring
high resolution imaging over wide fields.
Fast guiding, while offerering important resolution gains, will also
present its own challenges since 
one expects the PSF to become sytematically anisotropic
depending on location with respect to the guide stars
for the reasons described by \citeN{mfa+91}.
Also, fast guiding does not cure PSF anisotropies from telescope aberrations.

\subsection{Atmospheric Dispersion}

Variation of the refractive index of the atmosphere with wavelength will cause
images to be dispersed into spectra, and the consequent elongation of
images was estimated by KSB, who pointed out that this could potentially
cause problems when using PSF's of stars to correct the shapes of
galaxies, since the elongation depends on the
spectrum of the object; an object with a line-like spectrum,
or one with a sharp-edged absorption band which falls within
the bandpass, will of course be less dispersed than a continuum object.
Here we will make this more quantitative.

Weak-shear measurements are usually made on images composed from
multiple exposures taken at a range of air masses and which are
co-registered using centroids of foreground stars. The astrometric
solution obtained using stars of a variety of spectra will be a compromise.
Assuming, as
is usually the case, that one has many more stars at ones disposal
than the number of coefficients of the transformation one is
trying to solve for (typically a low order polynomial), then
the solution obtained by minimising least squared residuals will be
one which correctly maps surface brightness at some wavelength
$\lambda_0$ determined by the average properties of the foreground
stars, but with photons at other wavelengths displaced by an
amount proportional to $\lambda - \lambda_0$. The
sum of a set of $N$ such images is
\begineq
f_{\rm tot}(r, \lambda) =
\sum\limits_{I=1}^N f(r + \alpha_I (\lambda - \lambda_0), \lambda)
\endeq
where $f(r, \lambda)$ is the image seen at zenith.
The 2-vector $\alpha_{Ii}$ has
$|\alpha_I| = (\Theta_0 / \lambda_0) (\d \ln \Theta / \d \ln \lambda)
\tan z_I$, with $z_I$ the zenith
distance, and $\hat \alpha_I$ is directed towards the horizon.  
Tables of the
refraction angle $\Theta(\lambda)$ are given by \citeN{allen73}.
A star with SED $f_\lambda$, and therefore with photon distribution
function $S(\lambda) d \lambda \propto \lambda f_\lambda d \lambda$,
gives, in a photon counting system, a response at
zenith $f(r,\lambda) = \delta(r) S(\lambda)$, and the summed image
of such a star is then
\begineq
f_{\rm tot}(r) = \int d \lambda f_{\rm tot}(r, \lambda)
= \sum\limits_{I=1}^N\int d\lambda \delta(r - \alpha_I (\lambda - \lambda_0))
S(\lambda)
\endeq
The centroid of a star in the image $f_{\rm tot}(r)$ is an 
average of the centroids that
would be obtained from each of the contributing images.  Consider
the $I$'th image, and rotate the coordinate system so that $\alpha_I$
lies along the $x$-axis.  The centroid for this component is
$\overline r_I = ({\overline x}_I, 0)$ with
\begineq
{\overline x}_I = {
\int dx \int dy \int d\lambda \;
x \delta(x - \alpha_I(\lambda - \lambda_0)) \delta(y) S(\lambda)
\over
\int dx \int dy \int d\lambda \;
\delta(x - \alpha_I(\lambda - \lambda_0)) \delta(y) S(\lambda)
}
= {\alpha_I \int d\lambda\; (\lambda - \lambda_0) S(\lambda)
\over  \int d\lambda\; S(\lambda)}
= \alpha_I ({\overline \lambda} - \lambda_0)
\endeq
from which it follows that the $\lambda_0$ which minimises
$\langle \overline x_I^2 \rangle$, being
the average over the stars used for registration of the
squared displacement, is $\lambda_0 = \langle \overline \lambda \rangle
= {1 \over n_{\rm stars}} \sum \overline \lambda$, i.e.~simply the
average over the registration stars of their mean wavelength.
The centroid for the summed image is
\begineq
\overline r_i = ({\overline \lambda} - \lambda_0) \langle \alpha_i \rangle_I
\endeq
where $\langle \alpha_i \rangle_I \equiv {1 \over N} \sum \alpha_{Ii}$.
Similarly, the central second moment is
\begineq
\label{eq:pijdispersion}
p_{ij} = 
{\int d^2 r \; (r_i - \overline r_i)(r_j - \overline r_j)f_{\rm tot}(r)
\over
\int d^2 r 
f_{\rm tot}(r)}
 = \overline{(\lambda - \lambda_0)^2}
\langle \alpha_i \alpha_j \rangle_I
- (\overline \lambda - \lambda_0)^2 
\langle \alpha_i \rangle_I 
\langle \alpha_j \rangle_I
\endeq
Note that since, for any sensible zenith angle, the width of the spectrum
is tiny compared to that of the PSF, this unweighted second moment
fully characterises the effect of dispersion.

\begin{figure}[htbp!]
\centering\epsfig{file=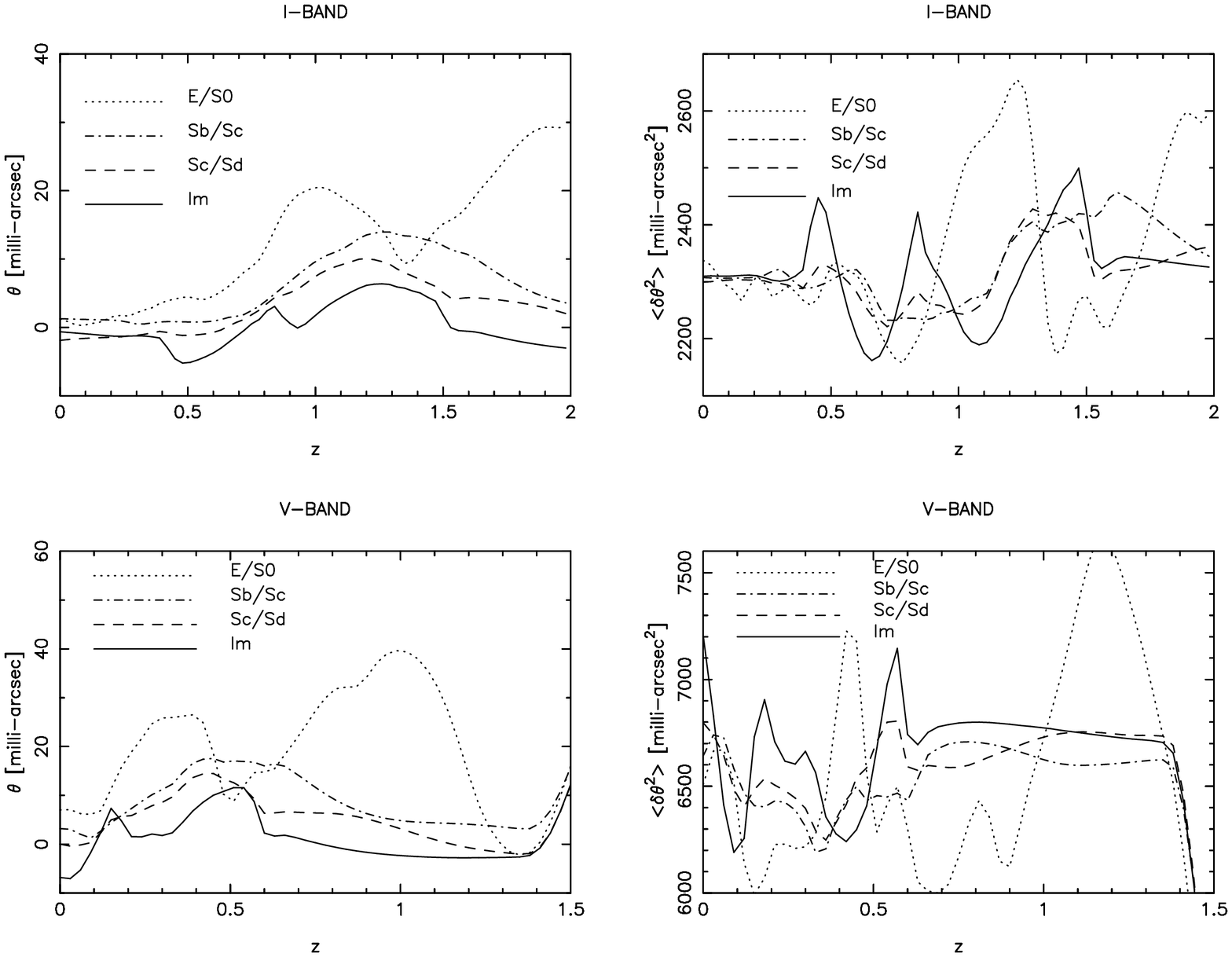,width={1.0 \linewidth},angle=0}
\figcap{Atmospheric dispersion. Left hand panels show the
angular displacement due to changes in the SED for galaxies
at a range of redshifts.  A zenith angle of 45 degrees was
assumed, and upper and lower panels show the deflection for
the I-band and V-band system response functions for the
CFH12K camera. 
More concretely, the quantity plotted is the mean
wavelength 
$\overline \lambda - \lambda_0 = \int d \lambda \; (\lambda - \lambda_0)
\lambda f_\lambda R(\lambda)
/ \int d \lambda \; \lambda f_\lambda R(\lambda)$ where 
$f_\lambda$ is the SED, $R(\lambda)$ the system response, 
$\lambda_0$ was taken to be $820$nm, and
$\overline \lambda - \lambda_0$ was
converted to angle as described in the text, and assuming
a 4000m observatory altitude.
The V-band plot has been limited to $z < 1.5$ since the
CWW SED's are limited to proper wavelength $\lambda > 200$nm and
redshift past the $V$-band at around $z = 1.5$.
The plots shows that the shifts due to SED variation are typically
on the order of $20$mas in the I-band, but somewhat larger in V.  
The elliptical/S0 SED shows a somewhat
greater excursion at high redshift, but this it caused by the
SED red-shifting right out of the filter band, so one
would not expect to find many such objects in flux limited samples.
The panels on the right show how the width of the
spectrum, and hence the PSF polarization, 
varies with redshift and galaxy type.  Here the
quantity plotted is 
$ \overline{(\lambda - \lambda_0)^2}
= \int d \lambda \; (\lambda - \lambda_0)^2 \lambda f_\lambda R(\lambda)
/  \int d \lambda \; \lambda f_\lambda R(\lambda)$,
again converted to angle, here assuming $\langle \tan^2 z \rangle = 1$.
}
\label{fig:dispersion}
\end{figure}

Consider, for illustration, the case of an equatorial field observed
with a telescope near the equator, and for a range of 
zenith angle $|z| < z_{\rm max}$.  In this case $\sum \alpha_I = 0$, so
the second term in (\ref{eq:pijdispersion}) 
vanishes (this is not a particularly unusual special case;
for a number of observations spread over a range of zenith
angles, one would generally expect the second term here to
become relatively small).  
When we solve for the PSF from the shapes of foreground stars of 
various types we are effectively
averaging over a mix of SED's appropriate for a low
redshift spiral galaxy. The average PSF moment obtained from the
stars can, for the equatorial case, then be written as
\begineq
\langle p_{xx} \rangle = 
(\langle \overline{\lambda^2} \rangle -
\langle \overline \lambda \rangle^2) \langle \alpha^2 \rangle_I
\endeq
where by $\langle \alpha^2 \rangle_I = 
(\Theta_0 / \lambda_0)^2 (\d \ln \Theta / \d \ln \lambda)^2
{1\over N} \sum_I \tan^2 z_I$, and the SED dependent
coefficient of $\langle \alpha^2 \rangle_I$ is the
mean of the dispersions of the stars plus the
dispersion of the means of the stars.  For the equatorial
case this is the same as the
second moment for a single point like object with an SED 
like the average SED of the registration stars.  This is not
strictly true in general, but the difference is probably
a minor one, and it then follows that 
if the faint galaxies have SED's such that their 
$\overline{(\lambda - \lambda_0)^2}$ 
differs systematically from
that for a low redshift spiral then the PSF correction will
be systematically in error.
Figure \ref{fig:dispersion} quantifies the importance of this
effect, by using redshifted galaxy SED's of various types from
\citeN{cww80}.  
The panel on the left shows the displacement of the centroid of
galaxies of the various indicated types as their spectra
are red-shifted, and the right hand panel shows the variation of the
second moment of the PSF.   We find that the I-band second moment is
very stable, with peak fluctuations $\delta p_{lm} \sim 300
{\rm mas}^2$ for these parameters, and with little systematic
difference from a low spiral redshift galaxy if we integrate
over a range of redshifts.  For 
illustration, if we take the systematic change in
polarisation to be say $\delta p_{lm} \sim 100 {\rm mas}^2$ and
a Gaussian profile galaxy, this corresponds to a shear of
$\gamma \simeq 1.38 \delta p_{lm} / {\rm FWHM}^2 \simeq 
6 \times 10^{-4} (0''.5 / {\rm FWHM})^2$.
In the V-band the polarisation fluctuations are
larger by a factor 2-3, but even then, and even for
marginally resolved objects in excellent seeing the effect is
at or below the sensitivity for current and near future
surveys.  The effect scales as $\langle \tan^2 z \rangle$ however, so 
observations at $z \gg 1 $ should be avoided.  This would also
become more of an issue with fast guiding, where we can hope
to resolve much smaller objects.  The effect is also dependent on the
details of the system response function; the CFH12K system having
a particularly broad I-band response which exacerbates the effect.

In the foregoing we assumed that the foreground stars used for
registration were numerous and well mixed.  With a finite number of
stars there will be some additional color dependent error in the registration
which, if uncorrected, would give rise to artificial field
distortion.  However, we have found from simulations
\hide{The relative shift of centroids of stars according to their colors
can also cause artificial distortion in the image registration.
At a zenith distance of 45 degrees (air-mass
of 1.414), for instance, and for standard broadband
colors we find that centroids shift by about 10mas
per magnitude in $V-I$ color, and }
using the USNOA
catalog as the astrometric reference system that this introduces
artificial shear at around the $10^{-4}$ level, which is
negligibly small for present and near future surveys.

\subsection{Aberrations}

Aberrations of the optical elements of the telescope can be a significant
contribution to the anisotropy of the PSF.
These can be analyzed in
the same manner as the wavefront deformation due to the atmosphere, but
with a couple of distinctive features:  First, for low-order `classical' 
aberrations where the phase error varies smoothly,
and for ground based observing conditions, if an aberration is an
important factor then
it's contribution to the OTF will be nearly achromatic.
This is because if there is  a smooth  variation of the wavefront error
(rather than a Gaussian random field with power at all scales 
as in atmospheric turbulence)
amounting to $N \gg 1 $ wavelengths, then the PSF will be very 
well approximated by its
geometric optics limit with shape defined
by the pattern of caustics (though for a narrow band filter,
the PSF would actually be found on close examination to be composed of
a set of speckle sized patches 
concentrated along the lines where the classical caustics form \cite{bu80}).
The wavefront deformation can
be measured directly from out of focus images \cite{rr93} so this contribution
to the PSF can be directly predicted.

\subsection{Diffraction Limited Seeing}

Truly diffraction limited seeing arises when the rms phase error
across the aperture (due to the atmosphere and/or aberration of the
optical elements of the telescope) is much less than unity.  In this case the optical
transfer function $\tilde g(k)$ is, to a good approximation, just the auto-correlation of the aperture
$A\oplus A$ at lag $\Delta r = k D \lambda / 2 \pi$, and
must therefore vanish for spatial frequencies $k > 2 \pi / (f \lambda)$,
where $f$ is the ratio of the aperture diameter to the
focal length.  
The log of the OTF becomes ill
defined as one approaches the diffraction limit.
From (\ref{eq:OTF2}) we see that
this cut-off is also present in the case of turbulence dominated
seeing, but it occurs at a high frequency where the the optical
transfer function has already become exponentially small due to
atmospheric effects, and has little impact.

In the absence of aberrations, the OTF for diffraction limited
seeing is real and non-negative. The OTF is symmetric under rotations
of 180 degrees, and so any quadrupole anisotropy of the PSF
can be anulled simply by re-convolving one's image with
a 90 degree rotated PSF.
For diffraction limited  seeing the size of the PSF scales as the
inverse of the wavelength, a fact which can be incorporated in
empirical or theoretical \cite{krist95} modeling of the PSF.

In the HST WFPC2, figure errors are not negligible.  The imaginary
part of the OTF is excited to the degree that re-convolution with
a 90-degree PSF still leaves non-negligible PSF anisotropy.
The phase errors are not large --- the telescope is nearly
diffraction limited --- so this means that the wavelength
dependence could be quite complicated.  The systematic error
arising from differences between faint galaxy and foreground
star SED's can be estimated much as we did for atmospheric
dispersion.

\subsection{Guiding Errors, Pixellisation, and Detector Effects}
\label{subsec:pixellisation}

So far we have considered the continuous distribution of
intensity on the focal plane $\fobs(r)$.  In real
detectors we sample the image with a grid of pixels.
The response of a pixel in not uniform and has been directly
measured for front-illuminated EEV devices by 
scanning a small spot of light across the CCD
(\citeNP{jdo93}, \citeNP{jdo94}).
They found very little `leakage' of electrons across pixel boundaries, but
according to the \citeN{krist95} this is a substantial effect for the
WFPC2 instrument on HST; a photon landing in one pixel has a 
non-negligible probability of being detected by a neighboring pixel
instead.  Such effects are expected, and found, to depend on wavelength.
The value of a pixel is a sample of the convolution of the sky surface 
brightness with the `pixel function' $p(r)$.

In many real systems the pixel spacing $d$ is not much less than the
instrumental resolution and
images from single exposures
are quite badly under-sampled.  For this and other reasons,
images are typically constructed from a series of exposures with
either systematically (in the case of HST) or
randomly (for terrestrial observations) staggered positions
on the sky.  Each image gives a 2-dimensional 
grid of samples of $p \otimes \fobs$, and a piecewise
continuous function $f_I$ can be constructed by shifting the
grid of delta-functions into an absolute astrometric coordinate system
and convolving with some interpolation function $p_{\rm interp}$.
One can incorporate the effect of guiding errors on a 
single exposure as a convolution with the pixel function. 
If we average a set of $N$ such images the result is
\begineq
F(r) = {1\over N}\sum_I [(\fobs \otimes p). c_{\Delta_I}] \otimes p_{\rm interp}
\endeq
where $c_\Delta(r)$ is a 2-dimensional comb function with
spatial offset (in units of the pixel spacing) $\Delta$:
\begineq
c_\Delta(r) = \sum\limits_{i_x, i_y = -\infty}^{\infty} \delta (r - (i + \Delta))
\endeq
The form of $p_{\rm interp}$ depends on the type of interpolation used.
For `nearest pixel' interpolation $p_{\rm interp}$ is just a
uniform box of side $d$, but if one linearly interpolates between the pixel samples,
for example, then $p_{\rm interp}$ will be a more extended, but again readily
computable, function.

The transform of $F(r)$ is
\begineq
\tilde F(k) = \tilde p_{\rm interp} 
\sum \limits_{m_x, m_y = -\infty}^{\infty} (\tilde \fobs \tilde p)_{k - 2 \pi m / d}
{1\over N} \sum_I \exp(2 \pi i m\cdot \Delta_I)
\endeq
The $m_x=m_y=0$ term in the first sum here is just the
ideal image $\fobs$ convolved with $p$ and with $p_{\rm interp}$, while the
higher order terms represent aliasing. 
The transform $\tilde F(k)$ being $p_{\rm interp}$ times the superposition of a grid of
images of $\tilde p \tilde \fobs$ with spacing $2 \pi / d$. Since
$\tilde p$ is a fairly compact function the dominant aliasing comes
from the low-order images $m = \pm 1$. 
Aliasing is most severe
for a single exposure since the low-order aliased images contribute with unit
weight.
If we average $N$ randomly shifted images then the
strength of the $m \ne 0$ terms is reduced by a factor $\sim 1 / \sqrt{N}$,
aliasing is greatly reduced, and to a good approximation
the field $F(r)$ is simply the convolution of $\fobs$ with
$p \otimes p_{\rm interp}$.  With systematically staggered
images, as is possible with HST and potentially with fast on-chip
guiding,
one can do even better; with a uniform $M \times M$ grid of offsets
covering the  unit pixel, the nearest, and therefore most
problematic, modes $m_x, m_y = \pm 1$ are then zero and the
modes remain small until we get to a multiple of $2 M$ times the
Nyquist frequency. 
This assumes that the transformation from detector to sky
coordinates is determined and applied accurately.  If we make
finite errors $\delta \Delta$ in registration, 
the resulting image will be the
convolution of the ideal PSF with
a highly compact cluster of delta-functions, and the optical transfer function
$\tilde g$ will be the product of the atmospheric, telescope
transfer functions with the Fourier transform of this pattern.
In practice, one can typically register images to a small
fraction of a pixel (say $\lsim 0.05$ pixels), 
and the effect of inaccuracy
at this level will have negligible effect on the final PSF.

Noise in the images, assumed to be incoherent Poisson noise in
the source images, can be analyzed in a similar manner and
we find that the two-point function of the noise is 
just the convolution of $p_{\rm interp}$ with itself, and the
two-point function of the noise in re-circularized images can
be obtained by convolving the raw noise ACF with $\gdag$ twice.

\section{Simulated Data}
\label{sec:simulation}

To test the procedures described here we have generated simulated mock
data and then analysed these.   The simulations were made to match
as closely as possible observations of $\sim 3$hr integration
on the CFHT with $0''.6$ seeing.

\begin{figure}[htbp!]
\centering\epsfig{file=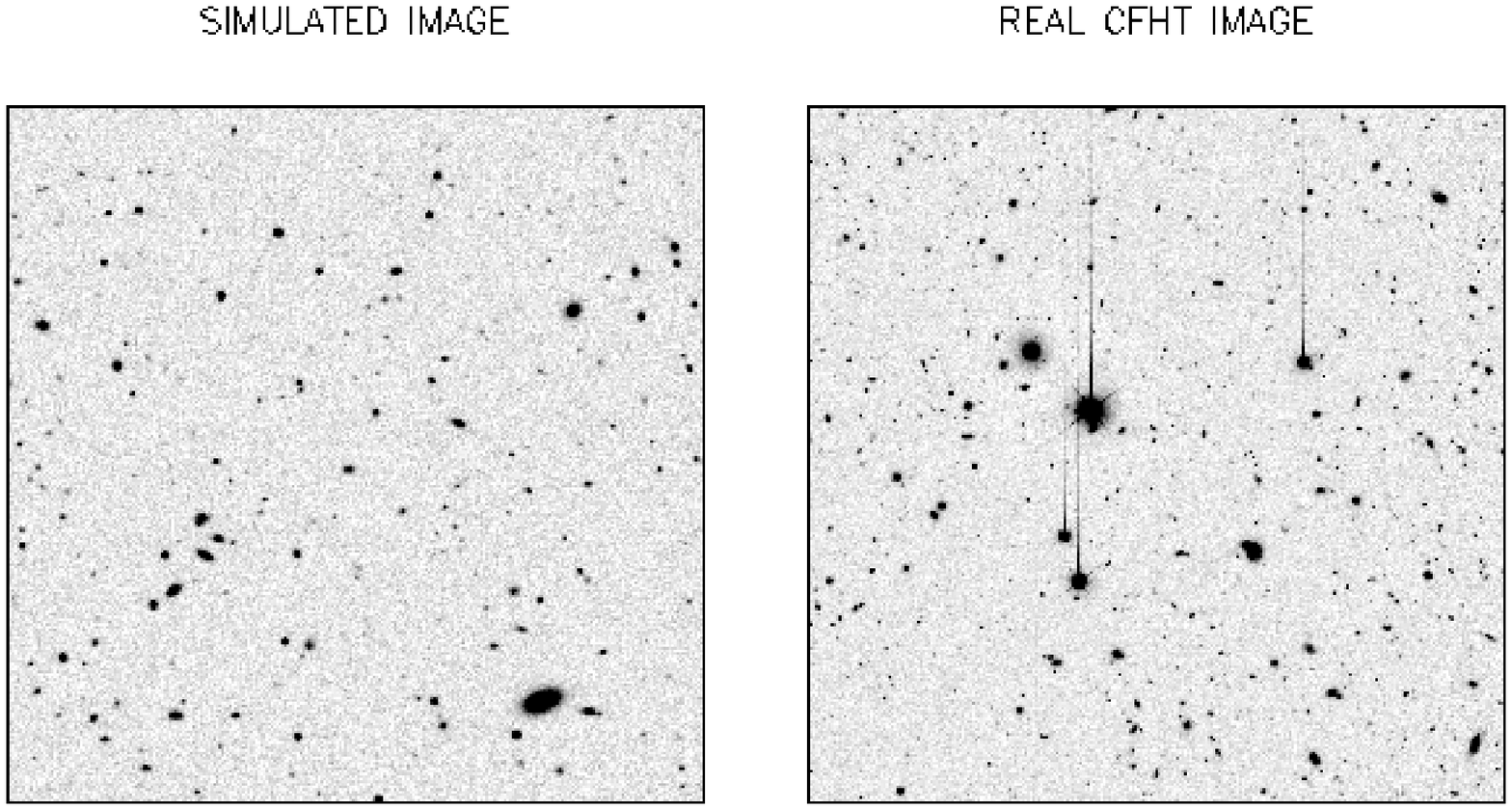,width={1.0 \linewidth},angle=0}
\figcap{Simulated and real image sections $2'.56$ on a side.}
\label{fig:cfhvssim_images}
\end{figure}

\begin{figure}[htbp!]
\centering\epsfig{file=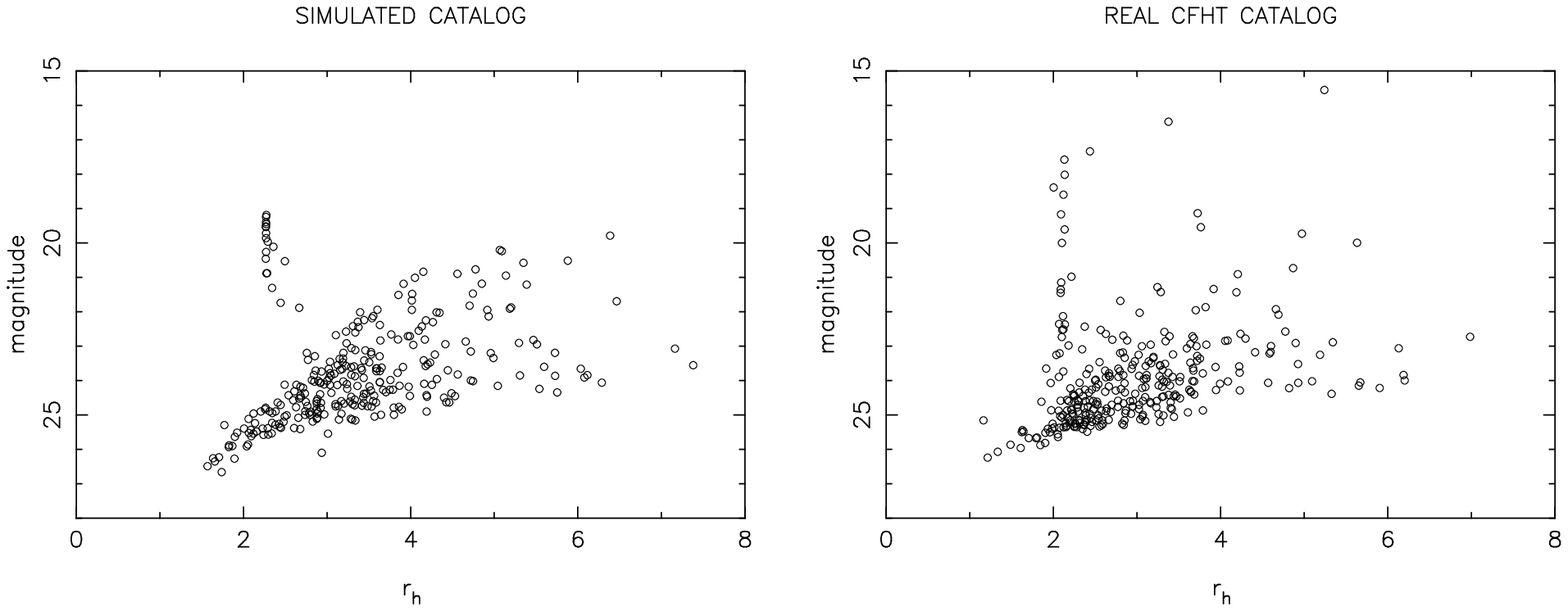,width={1.0 \linewidth},angle=0}
\figcap{Catalogs from image sections shown in figure \ref{fig:cfhvssim_images}.
}
\label{fig:cfhvssim_cats}
\end{figure}

We first generated a set of 200 mock catalogue of galaxies each
corresponding to a patch of sky of size $2'.56$ on a side.  Galaxies were 
drawn from a Schechter style luminosity function laid down in
a Poissonian manner in an Einstein de Sitter cosmology.  Images with pixel scale
$0''.075$ were then generated by realising the galaxies as
exponential disks
with
random orientations and scale lengths corresponding
to fixed rest frame central surface brightness.  The galaxies were modelled as
optically thick, since the optically thin model looks unrealistic
as it has too many very bright edge-on systems as compared to real images.
A number of point-like stars were added to the images, which were then
sheared with $\gamma = 0.1$, convolved with a 
Kolmogorov turbulence PSF with $0''.6$ FWHM, and then rebinned to $0''.15$ pixel 
scale.  When the real data are analysed they are interpolated from the
original $0''.2$ pixel scale to the final $0''.15$ scale with bi-linear
interpolation.  This results in a further convolution of the signal and
the noise in the real images, but with slightly different smoothing
kernels.  These kernels were computed by modelling the image shifts as
a uniform distribution within the final pixel size; the noise-free mock
images were convolved with the appropriate kernel and then Gaussian
white-noise images were generated to model the sky noise, and were convolved
with the appropriate kernel and then added to the images. A sample
image is shown in figure \ref{fig:cfhvssim_images}, and the 
corresponding size-magnitude diagram is shown in figure
\ref{fig:cfhvssim_cats}, from which it is apparent
that the simulated objects have properties very similar
to those detected in the real data.

These data were analysed exactly like the real data.   That is,
the objects were detected as peaks of a smoothed image.  The stars were
extracted and their shapes fit in the manner described to obtain the PSF.
A smoothed image $f_s = g \otimes \fobs$ was generated and from this the polarisation
$q_\alpha$ was computed using (\ref{eq:polarisationfromfs}), 
and the polarisability for each
object was computed using (\ref{eq:Pab3}).

\end{document}